\documentclass[twocolumn,english,prb,showpacs,superscriptaddress,a4paper]{revtex4-1}
\usepackage[T1]{fontenc}
\usepackage[latin9]{inputenc}
\usepackage{babel}
\usepackage{verbatim}
\usepackage{amsmath}
\usepackage{amssymb}
\usepackage{graphicx}
\usepackage{esint}
\PassOptionsToPackage{version=3}{mhchem}
\usepackage{mhchem}

\usepackage{array}
\usepackage{multirow}
\usepackage{bigstrut}
\usepackage{bigdelim}
\usepackage{tabularx}

\makeatletter
\@ifundefined{textcolor}{}
{%
 \definecolor{BLACK}{gray}{0}
 \definecolor{WHITE}{gray}{1}
 \definecolor{RED}{rgb}{1,0,0}
 \definecolor{GREEN}{rgb}{0,1,0}
 \definecolor{BLUE}{rgb}{0,0,1}
 \definecolor{CYAN}{cmyk}{1,0,0,0}
 \definecolor{MAGENTA}{cmyk}{0,1,0,0}
 \definecolor{YELLOW}{cmyk}{0,0,1,0}
}

\usepackage{amsfonts}\@ifundefined{definecolor}
\usepackage{color}{}
\usepackage{bm}
\usepackage{multirow}
\usepackage{bigstrut}

\newcommand{\bfr}{{\bf r}}
\newcommand{\be}{\begin{eqnarray}}
\newcommand{\ee}{\end{eqnarray}}
\newcommand{\ba}{\begin{array}}
\newcommand{\ea}{\end{array}}
\newcommand{\no}{\nonumber}

\begin{document}

\global\long\def\l{L}
\global\long\def\fb{\bar{f}}
\global\long\def\fc{f^{*}}
\global\long\def\fbc{\bar{f}^{*}}
\global\long\def\np{\nabla_{\!{\scriptscriptstyle \!+}}}
\global\long\def\nm{\nabla_{\!\!{\scriptscriptstyle -}}}
\global\long\def\K{\mathcal{K}}
\global\long\def\bperp{B_{\perp}}
\global\long\def\bpar{B_{\parallel}}
\global\long\def\e{\mathcal{E}}
\global\long\def\mymu{z}
\global\long\def\li{\text{Li}}

\makeatother

\title{Fluctuation conductivity of disordered superconductors in magnetic fields}
\author{Brian Tarasinski}
\email{tarasinski@lorentz.leidenuniv.nl}
\affiliation{Dahlem Center for Complex Quantum Systems and Institut f\"ur Theoretische
Physik, Freie Universit\"at Berlin, 14195 Berlin, Germany}
\affiliation{Instituut-Lorentz, Universiteit Leiden, P.O. Box 9506, 2300 RA Leiden, The Netherlands}
\selectlanguage{english}

\author{Georg Schwiete}

\affiliation{Dahlem Center for Complex Quantum Systems and Institut f\"ur Theoretische
Physik, Freie Universit\"at Berlin, 14195 Berlin, Germany}

\date{\today}
\begin{abstract}
We calculate fluctuation corrections to the longitudinal conductivity of disordered superconductors subject to an external magnetic field. We derive analytic expressions that are valid in the entire metallic part of the temperature-magnetic field phase diagram as long as the effect of the magnetic field on the spin degrees of freedom of the electrons may be neglected. Our calculations are based on a kinetic equation approach. For the special case of superconducting films and wires in parallel magnetic fields we perform a detailed comparison with results that were previously obtained with diagrammatic perturbation theory in the imaginary time formalism. As an application, we study the fluctuation 
conductivity of films in tilted magnetic fields with a special focus on the low-temperature regime. We present a detailed discussion of the phenomenon of the non-monotonic magnetoresistance and find that it displays a pronounced dependence on the tilting angle. 
\end{abstract}
\pacs{74.40.-n, 74.25.fc}

\maketitle

\section{Introduction}

The theory of superconducting fluctuations has been the subject of intense study
for many years.\cite{Larkin04} In the metallic part of the phase diagram, outside
the superconducting regime, Cooper pairs may form for a finite time. The
presence of these fluctuating Cooper pairs affects both thermodynamic and
transport properties of the metal. The phase transition between the metallic and
the superconducting state may be tuned by temperature or by so-called
pair-breaking mechanisms which lead to a partial or even complete destruction of superconductivity.\cite{Maki69} Examples of pair-breakers include magnetic
impurities in s-wave superconductors, external magnetic fields or a flux
penetrating a superconductor with doubly connected geometry. Fluctuation effects
are particularly strong for low-dimensional superconductors and further enhanced
by the presence of impurities.\cite{Larkin04} Detailed experimental\cite{Liulong01,Pourret06,Steiner06,Koshnicklong07,Sacepe10,Sternfeld11,Baturina12,Breznay12}  and theoretical studies\cite{Galitski01,Lopatin05,Shah07,Schwiete09,Schwiete10b,DelMaestro09,Michaeli09,Serbyn09,Glatz11a,Glatz11b,Khodas12,Tikhonov12,Petkovic13} of fluctuation phenomena in superconductors have become available in recent years. 

The subject of this paper is the calculation of fluctuation corrections to
conductivity in the metallic phase of disordered superconductors. The origin of
this field dates back to the work of Azlamazov and Larkin.\cite{Aslamazov68} These authors studied
the direct contribution of fluctuating Cooper pairs to conductivity close to the
transition temperature $T_{c0}$, the so-called paraconductivity. Shortly afterwards, additional contributions were discovered.\cite{Maki68,Thompson70} This development went
hand in hand with the study of different classes of diagrams in many-body
perturbation theory. It became customary to divide the set of most relevant
diagrams into three classes, the Aslamazov-Larkin diagram and the density of
states and Maki-Thompson diagrams.\cite{Larkin04}  

Initially, studies focused around the vicinity of $T_{c0}$ for vanishing or small
magnetic fields. In Ref.~\onlinecite{Galitski01}, the fluctuation conductivity was calculated for disordered superconducting films in perpendicular magnetic fields in the vicinity of the critical magnetic field $B_{c2}$. It was established that at
very low temperatures superconducting fluctuations lead to a
non-monotonic magnetoresistance (NM); close to $B_{c2}$, the resistance
curve displays a maximum as a function of the magnetic field. In another
theoretical study, Ref.~\onlinecite{Lopatin05,Shah07}, it was found that the NM also exists in the vicinity of certain other classes of pair-breaking transitions such as for films and wires in parallel magnetic fields. 

Recently, a novel scheme for deriving fluctuation conductivity was introduced,\cite{Tikhonov12}
which is based on the Usadel equation.\cite{Usadel70} The calculation is performed in the
Keldysh formalism to circumvent the analytic continuation necessary in the Kubo
technique.\cite{AGD63} For temperatures close to $T_{c0}$ and in the absence of a magnetic
field, the Usadel equation has been used for the calculation of fluctuation
conductivity before.\cite{Volkov98} In Ref.~\onlinecite{Tikhonov12}, in turn, general expressions for the
fluctuation conductivity in disordered superconducting films with perpendicular
magnetic field were derived for the whole normal part of the temperature-magnetic field phase diagram
(outside the strong fluctuation regime close to the transition line). In this
approach, it was possible to identify three distinct contributions to
conductivity at the very early stages of the calculation. The first one, termed
density of states correction ($\delta \sigma_{dos}$), is seen to be directly related to the
change in the quasiparticle density of states. The second contribution is the
anomalous Maki-Thompson 
correction ($\delta\sigma_{an}$), which is known from diagrammatic perturbation theory and describes a coherent
rescattering in the Cooper channel. The third term may be interpreted as the
direct contribution of Cooper pairs to the current, and was therefore named
supercurrent correction ($\delta \sigma_{sc}$). It should be noted that the density of states and supercurrent 
contributions in the Usadel equation approach are in general not identical to
the contributions of the density of states and Aslamazov-Larkin diagrams in the
conventional classification. 

Let us briefly recall the origin of the NM for the perpendicular magnetic field
case using the language introduced in Ref.~\onlinecite{Tikhonov12}. We discuss the low temperature
regime $t=T/T_{c0}\ll 1$ close to the (temperature-dependent) critical field
$B_{c2}(T)$, so that $h=(B-B_{c2}(T))/B_{c2}(T)\ll 1$. For $t\gg h$, all
corrections, $\delta\sigma_{dos}$, $\delta \sigma_{an}$, and $\delta \sigma_{sc}$, contribute, and the total correction to
conductivity is positive. As one moves further away from the transition line at
fixed temperature, in the limit $t\ll h$, the anomalous Maki-Thompson correction becomes ineffective. The density of states correction to conductivity, which is negative, and the supercurrent currection, which is positive, are of similar magnitude. The density of states correction dominates,  however, leading
to a net negative correction to conductivity. For large magnetic fields, the
negative density of states correction is still dominant, but eventually diminishes. The result
is a non-monotonic magnetoresistance. Close to $B_{c2}$, the results of
Ref.~\onlinecite{Tikhonov12} coincide 
with those obtained by Galitski and Larkin,\cite{Galitski01} who specifically focused on this
regime and used the conventional diagrammatic method for the calculation. It
should be noted that the low-temperature regime is quite different from the
well-studied case of small magnetic fields for $T\approx T_{c0}$. In the latter
regime, $\delta \sigma_{sc}$ is much larger in magnitude than $\delta \sigma_{dos}$. The main
difference is that close to $B_{c2}$ Landau level quantization of the Cooper
pair propagator becomes crucial. This is why the supercurrent correction becomes less
singular and $\delta \sigma_{sc}$ and $\delta \sigma_{dos}$ are of a comparable magnitude.

In this paper, we use the Usadel equation approach to derive general expressions
for the fluctuation conductivity in superconductors subject to a magnetic field.
We assume that the sample geometry is translationally invariant along the
direction of the electric field, while the sample may be confined in the
transverse direction(s). The derived formulas are in particular applicable for
superconducting wires, superconducting films in magnetic fields of arbitrary
orientation and for cylinders threaded by a magnetic flux. As a specific
application, we study in detail the phenomenon of the NM for films in tilted
magnetic fields. We focus on the low temperature regime, and describe the
evolution of the NM as a function of the tilting
angle. While the phenomenon persists for any angle, there are two distinct
regions, one comprising the parallel magnetic field case and the other one the
perpendicular magnetic field case, for which the physical origin of the
phenomenon as well as the 
magnitude of the resulting resistance-maximum are quite distinct. The cross-over
between the two regimes occurs for almost parallel magnetic fields.

The theory developed in this paper is applicable in the limit of weak disorder, $\epsilon_F\tau\gg 1$, where $\epsilon_F$ is the Fermi energy and $\tau$ the transport scattering time. From the experimental perspective, detailed low-temperature resistance measurements have been performed on weakly disordered films in perpendicular magnetic fields, see, e.g., Refs. \onlinecite{Gantmakher03}, \onlinecite{Baturina05} and \onlinecite{Steiner06}. Measurements on films in parallel\cite{Gantmakher00,Parendo04} and tilted\cite{Johansson11} magnetic fields exist, but focused on more strongly disordered films in the context of the so-called superconductor-insulator transition. 

For the case of superconductors in parallel magnetic fields, we perform a detailed comparison of our results to those obtained
in Ref.~\onlinecite{Lopatin05,Shah07} with the help of the traditional diagrammatic technique. We show that
there is a one-to one correspondence between the results obtained in the two
formalisms (up to details of the ultraviolet regularization). The mapping is not
simple, however. The three distinct contributions identified in the kinetic
equation approach correspond to a mixture of terms originating from different
diagrams. This comparison is motivated by a discrepancy between recent results reported
for films in perpendicular magnetic fields. Glatz, Varlamov and Vinokur \cite{Glatz11a,Glatz11b}
used the traditional approach in the imaginary time formalism for the
calculation of the fluctuation conductivity. The results of this study disagree
with a number of previously reported results, including those for $B\approx
B_{c2}$, Ref.~\onlinecite{Galitski01}, and the high temperature regime $T\gg T_{c0}$ at $B=0$, Ref.~\onlinecite{Altshuler83}. The
technically very different Usadel equation approach of Ref.~\onlinecite{Tikhonov12}, however, confirmed
these earlier results. The comparison performed here for the parallel magnetic field case demonstrates an agreement between the imaginary time
formalism as worked out in Ref.~\onlinecite{Lopatin05,Shah07} and the Usadel equation approach on the level
of general formulas valid in the entire normal part of the phase diagram. This
includes, in particular, the high temperature regime $T\gg T_{c0}$ for $B=0$, which
lies within the range of applicability of all the mentioned works. 

The paper is organized as follows. In Sec.~\ref{sec:results} we present the main
results of our study. Specifically, in
Sec.~\ref{subsec:genresults}, we display the general formulas for the fluctuation
conductivity in superconductors subject to a magnetic field and discuss their
range of applicability. In Sec.~\ref{subsec:tilted}, we discuss the
fluctuation conductivity of a thin amorphous superconducting film in a tilted
magnetic field.  The example of the film in a tilted magnetic field constitutes
a special application of the general formalism outlined in this paper.  The remainder of the paper is devoted to the technical
details of the approach as well as to a comparison with the traditional
diagrammatic technique. In Sec.~\ref{sec:general} we introduce the Usadel equation
approach underlying the calculation of fluctuation conductivity as well as the
derivation of the results presented in Sec.~\ref{sec:results}. The formalism we use is
a generalization of the approach introduced in Ref.~\onlinecite{Tikhonov12} so as to include
pair-breaking effects. We 
outline the main steps of the derivation in a condensed form in order to make
the paper self-contained. In Sec.~\ref{sec:comparison} we specialize to the parallel magnetic field case and compare our results to
those obtained in Ref.~\onlinecite{Lopatin05,Shah07} using the traditional diagrammatic approach. The
results of the comparison are summarized in table \ref{tab:table}. Sec.~\ref{sec:tilted}
is devoted to films in a tilted magnetic field. Here, we present the derivation
of the results presented in Sec.~\ref{subsec:tilted}. Eventually, in Sec.~\ref{sec:conclusion} we
conclude. Some technical details of the calculation are relegated to two
appendices.

\section{Results}
\label{sec:results}
In this section, we present the main results of our study. We split
the discussion into two parts. In the first part, we present the results
for the general theory of fluctuation transport in superconductors subject to a magnetic field. 
In the second part, we discuss the fluctuation conductivity of a thin disordered film
in a tilted magnetic field. The presentation is intended to be self-contained. Details of the derivation are described in Secs.~\ref{sec:general} and \ref{sec:tilted}.

\subsection{General Results: Disordered superconductors in a magnetic field}
\label{subsec:genresults}
The main result of this paper are expressions for the fluctuation conductivity in disordered superconductors subject to a magnetic field. In Sec.~\ref{subsec:tilted} below, we discuss the case of a thin film in a tilted magnetic field as an application. The general results, however, are applicable not only to thin films, but also to several other geometries, for example wires, cylindric tubes, and nanoribbons. The difference between these examples lies in the spectrum of superconducting fluctuations. In the following, we present the equation determining the fluctuation spectrum, Eq.~\eqref{eq:eigenvalue-equation-1}, and write down the general results for the corrections to
conductivity, Eqs.~\eqref{eq:dosansc} to \eqref{eq:general-al2b-1}. 
Then, we briefly discuss the range of applicability.

\subsubsection{Fluctuation spectrum}

The fluctuation propagator of the superconducting order parameter field, cf. Eqs.~\eqref{eq:L} and \eqref{eq:E} below, is diagonal in the basis of eigenfunctions determined by the following eigenvalue equation
\begin{equation}
	-\frac{D}{2}[\mathbf{\nabla} - 2ie\mathbf{A}({\bf r})]^{2}\phi_{n}\left({\bf r}\right)=\alpha_{n}\phi_{n}\left({\bf r}\right).\label{eq:eigenvalue-equation-1}
\end{equation}
This equation is similar to the single-particle Schr\"odinger equation in quantum mechanics. Here, however, it is related to the motion of Cooper pairs. Due to the diffusive nature of Cooperons, the mass entering the conventional Schr\"odinger equation is replaced by the inverse of the diffusion constant $1/D$. The solutions of this equation depend on the external magnetic field and on the geometry of the system, as the equation needs to be supplemented with appropriate boundary conditions. For the boundary to an insulator or vacuum, the following condition should be chosen 
\begin{equation}
{\bf n}\cdot (\mathbf{\nabla} - 2ie\mathbf{A})\phi =0,
\end{equation}
where ${\bf n}$ is the vector normal to the boundary. This condition corresponds to the requirement of zero super-current through
the boundary.

The information about the eigenfunctions $\phi_n$ and corresponding eigenvalues $\alpha_n$ for a certain geometry is sufficient in order to obtain the fluctuation corrections.

\subsubsection{General expressions for fluctuation corrections}

We now present the results for the fluctuation corrections (in this section and the rest of this manuscript, we set $\hbar=1$). We write the total correction as the sum of three parts,
\be
\delta \sigma=\delta\sigma_{dos}+\delta\sigma_{an}+\delta \sigma_{sc},\label{eq:dosansc}
\ee
corresponding to the classification in the Usadel equation scheme. The density of states contribution takes the form
\be
&&\delta \sigma_{dos}\label{eq:dos}\\
&=&2De^{2}\int\frac{d\omega}{2\pi}\sum_{n}\rho_{nn}\left[\mathcal{B}'\mbox{Re}\e_{n}'\mbox{Im}\l_{n}-\mathcal{B}\mbox{Im}\left(\e_{n}^{\prime\prime}\l_{n}\right)\right]\no.
\ee
This correction originates from the suppression of the quasiparticle density of states near the Fermi surface.

The anomalous Maki-Thompson correction reads
\be
	\delta \sigma_{an}=2De^{2}\int\frac{d\omega}{2\pi}\sum_{n}\rho_{nn}\frac{\mathcal{B}'}{\alpha_{n}}\mbox{Im}\l_{n}\mbox{Im}\e_{n}\label{eq:general-mt-1}.
\ee
This correction may be interpreted as a resonantly enhanced interference effect in the Cooper channel.

The correction induced by the fluctuating supercurrent is conveniently written as the sum of three terms
\be
\delta \sigma_{sc}=\delta\sigma_{sc}^{(1)}+\delta\sigma_{sc}^{(2a)}+\delta\sigma_{sc}^{(2b)},
\ee
where
\be
\delta \sigma_{sc}^{\left(1\right)} &=&-De^{2}\int\frac{d\omega}{2\pi}\sum_{nm}\frac{\boldsymbol{d}_{nm}\mathcal{B}}{\alpha_{n}-\alpha_{m}}\left(\e_{n}^{\prime}-\e_{m}^{\prime}\right)\left(\l_{n}-\l_{m}\right),\no\\
&&\label{eq:general-al1-1}
\ee
and
\be
\delta \sigma_{sc}^{\left(2a\right)} &=&-De^{2}\int\frac{d\omega}{2\pi}\sum_{nm}\frac{\boldsymbol{d}_{nm}\mathcal{B}'}{\alpha_{n}-\alpha_{m}}\mbox{Im}\left(\e_{n}-\e_{m}\right)\label{eq:general-al2a-1}\\
&&\qquad \times \mbox{Im}\left(\l_{n}-\l_{m}\right),\no \\
\delta \sigma_{sc}^{\left(2b\right)} &=&De^{2}\int\frac{d\omega}{2\pi}\sum_{nm}\frac{\boldsymbol{d}_{nm}\mathcal{B}'}{\alpha_{n}-\alpha_{m}}\mbox{Re}\left(\e_{n}-\e_{m}\right)\label{eq:general-al2b-1}\\
 && \qquad\times \mbox{Re}\left(\l_{n}-\l_{m}+\left(\e_{n}^{*}-\e_{m}\right)\l_{n}\l_{m}^{*}\right).\nonumber
\ee

We introduced the retarded fluctuation propagator in equilibrium
\begin{equation}
L_n(\omega)=\frac{1}{\e_n(\omega)},\label{eq:L}
\end{equation}
where
\begin{equation}
\e_{n}\left(\omega\right)=\ln\frac{T_{c0}}{T}+\psi\left(\frac{1}{2}\right)-\psi\left(\frac{1}{2}+\frac{2\alpha_{n}-i\omega}{4\pi T}\right),\label{eq:E}
\end{equation}
and $\psi$ denotes the digamma function.\cite{Abramowitz72} $\mathcal{B}\left(\omega\right)=\coth\left(\omega/2T\right)$ is the bosonic
equilibrium distribution function. The prime in the above set of formulas denotes a derivative with respect to frequency, $f'(\omega)=\partial_\omega f(\omega)$.
We further introduced the following matrix elements in the basis of eigenfunctions:
\begin{align}
\rho_{nm}\left({\bf r}\right) & =\phi_{n}^{*}\left({\bf r}\right)\phi_{m}\left({\bf r}\right)\label{eq:rho}\\
\boldsymbol{d}_{nm}\left({\bf r}\right) &
=\frac{1}{2}\hat{\bf E}\cdot{\bf r}_{nm}
\big(  \phi_{n}\left({\bf r}\right)[\nabla+2ie{\bf A}({\bf r})]\phi_{m}^{*}\left({\bf r}\right)\no\\ 
& \qquad -\phi_{m}^{*}\left({\bf r}\right)[\nabla-2ie{\bf A}({\bf r})]\phi_{n}\left({\bf r}\right) \big) 
\label{eq:d}
\end{align}
where $\hat{\bf E}$ is the unit vector in the direction of the external electric field and we further defined
\begin{equation}
	{\bf r}_{nm}=\int d{\bf r}\, {\bf r}\phi_{n}^{*}\left({\bf r}\right)\phi_{m}\left({\bf r}\right). \label{eq:rnm}
\end{equation}
From the form of the expressions for $\delta\sigma_{sc}$, Eqs.~\eqref{eq:general-al1-1} to \eqref{eq:general-al2b-1} and using the relation
$\boldsymbol{d}_{nm}=-\boldsymbol{d}^{*}_{mn}$, it follows that only the real
part of $\boldsymbol{d}_{nm}$ contributes, which turns out to be sufficient for the
longitudinal conductivity we study here.  When considering the transversal
conductivity, where $\boldsymbol{d}_{nm}$ is purely imaginary, a particle-hole
symmetry breaking term needs to be added to $\l$ for a non-zero result, and the
formulas given above do not hold. \cite{Tikhonov12}

We note that the index $n$, used in general to enumerate the eigensystem of Eq.~\eqref{eq:eigenvalue-equation-1}, 
might in fact be a multi-index with several components. It is also possible
that $n$ does not enumerate a discrete set, but rather a continuum. In that
case, the sum over $n$ has to be replaced by the corresponding integral. 

An important remark is in order here. The anomalous Maki-Thompson correction diverges in the absence of a magnetic field, as then $\alpha_0\rightarrow 0$. The correction may be regularized by introducing a finite dephasing rate $1/\tau_\phi$.\cite{Larkin04} Dephasing can be provided by magnetic impurities, electron-electron or electron-phonon collisions. For low temperatures, electron-electron collisions dominate. Outside the region of
strong fluctuations, one can consider the dephasing rate as energy-independent and equal to the sum of rates due to the Coulomb\cite{Altshuler82} and Cooper
channels.\cite{Larkin72a,Brenig85} In our study, we will treat $1/\tau_\phi$ as a phenomenological parameter; it may be be introduced into the theory by replacing $\alpha_n\rightarrow \alpha_n+1/2\tau_\phi$ in the formulas for $\delta \sigma$ and $\e_n$ given above. 

In principle, the equations presented above can be used to obtain the fluctuation
corrections to conductivity for any sample along an unconfined direction,
by solving the eigenvalue problem \eqref{eq:eigenvalue-equation-1}
for the given geometry. In the next section, we briefly list a number of cases, for which these formulas can be applied. 

\subsubsection{Eigenvalues $\alpha_n$ for different geometries}

In a bulk sample, in the absence of a magnetic field, the eigenvalue equation \eqref{eq:eigenvalue-equation-1} can be solved by Fourier transformation due to translational invariance, resulting in the following eigenfunctions
and eigenvalues: 
\begin{equation}
\phi_{q}\left({\bf r}\right)=e^{i{\bf q}\cdot {\bf r}},\quad\alpha_{q}=\frac{1}{2}Dq^{2}.
\end{equation}

Films and nanowires have a reduced dimensionality. For the unconfined
directions, it is useful to introduce a continuous Fourier transformation, while in the transverse
direction(s), modes are quantized. This remains true if a parallel
field is applied, because it is possible to introduce a vector potential that depends only on the transverse coordinate(s). The eigenvalues 
\begin{equation}
\alpha_{qn}=\frac{1}{2}Dq^{2}+\alpha_{\perp n}\label{eq:alpha-pair-breaking-1}
\end{equation}
can be written as a sum of a continuous and a discrete component.

If the transverse direction is small in extent compared to the superconducting
coherence length $\xi_{0}$, often only the lowest transverse mode
is relevant. The lowest eigenvalue $\alpha_{\perp0}$ then plays the
role of a pair-breaking parameter.\cite{Maki69,Lopatin05,Shah07} For example, for a film
in a parallel magnetic field one finds 
\begin{equation}
\alpha_{\perp0}=\frac{1}{6}D\left(eB\right)^{2}d^{2},
\end{equation}
where $d$ is the thickness of the film.

For a film with a \emph{perpendicular} field, the situation is different from the one discussed above, since the vector potential ${\bf A}$ cannot be chosen to be translationally invariant within the plane. The eigenvalues are degenerate Landau levels
\begin{equation}
\alpha_{n}=2eDB\left(n+\frac{1}{2}\right),
\end{equation}
which should be supplemented with suitably chosen eigenfunctions.\cite{Cohen80} For this case, the fluctuation conductivity has been discussed with the help of the Usadel equation approach in Ref.~\onlinecite{Tikhonov12}.

In this paper, we show that in a tilted field with both perpendicular and parallel field components, the relevant eigenvalues $\alpha_n$ can be written as a sum of the eigenvalues for the perpendicular and parallel magnetic field cases. We will discuss the fluctuation corrections for this case in detail.

Another interesting quasi one-dimensional system is a cylindric shell, i.e., a nanowire with annular cross-section.\cite{Liulong01, Sternfeld11} Here, a parallel magnetic field also gives rise to a flux threading the cylinder. The dependence of the lowest eigenvalue $\alpha_{\perp 0}$ on the flux is then (in the limit of vanishing thickness) periodic with the superconducting flux quantum $\varphi_0^{*}=h/2e$, as can be seen from solving the eigenproblem \eqref{eq:eigenvalue-equation-1}. This special case will be discussed in a separate publication.\cite{Tarasinski13a}

\subsection{Film in a tilted magnetic field}
\label{subsec:tilted}

\begin{figure}[tb]
\includegraphics[width=\columnwidth]{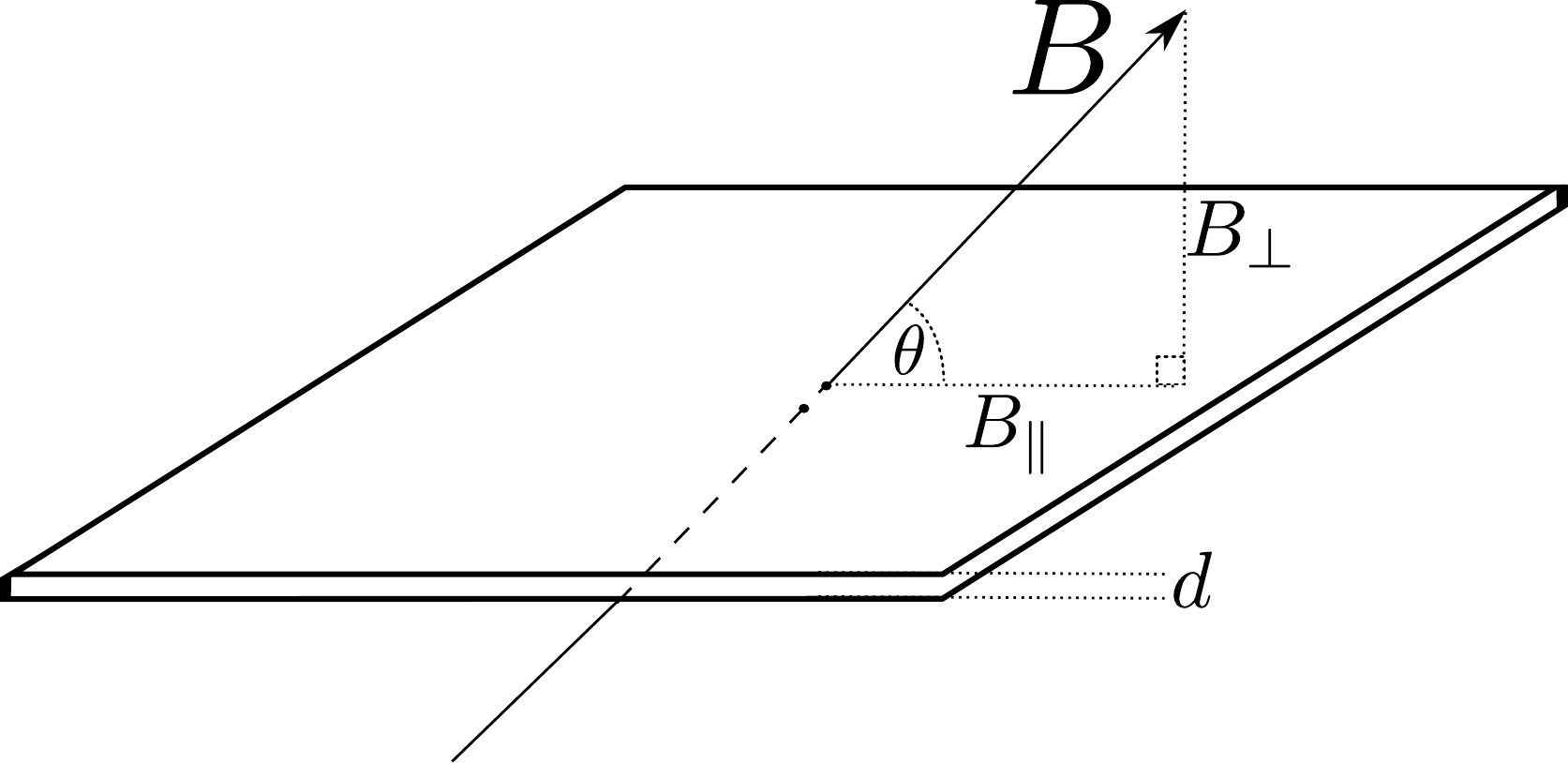}
\caption{A sketch of the system under study. A superconducting film of thickness $d$ is penetrated by a magnetic field of magnitude $B$ at an angle
$\theta$.}
\label{fig:A-sketch-of-the-system-1} 
\end{figure}

We will now discuss in detail the fluctuation corrections for a thin film in a tilted magnetic field.
We consider a thin amorphous superconducting
film of thickness $d$ penetrated by a magnetic field at an angle $\theta$, $0\leq\theta\leq\pi/2$,
measured between the field lines and the sample. We study the dirty
limit, i.e., $T_{c0}\tau\ll1$, where $T_{c0}$ is the critical
temperature of the superconductor and $\tau$ is the elastic scattering
time of the electrons. We choose coordinates so that the film lies
in the $x-y$ place, and the magnetic field ${\bf B}$ can be written
as (see Fig. \ref{fig:A-sketch-of-the-system-1}) 
\begin{equation}
{\bf B}=B\sin\theta\boldsymbol{\hat{z}}+B\cos\theta\boldsymbol{\hat{y}}.
\end{equation}
Here, $\boldsymbol{\hat{z}}$ and $\boldsymbol{\hat{y}}$ are unit vectors in the $z$ and $y$ direction, respectively.
We will sometimes use the notation $B_{\parallel}=B\cos(\theta)$
and $B_{\perp}=B\sin(\theta)$.

The film is assumed to be sufficiently thin so that the condition
$d\ll\xi_{0}$ is fulfilled, where $\xi_{0} \approx 0.36\sqrt{D/T_{c0}}$ is the superconducting
coherence length at zero temperature.\cite{Larkin04}
In this limit, the film can
be considered as two-dimensional as far as its superconducting properties
are concerned, whereas the electron motion is assumed to be three-dimensional.
In the following we neglect the destructive effect
on superconductivity caused by the direct coupling of the magnetic
field to the magnetic moment of the electrons. It is known that this is a good approximation
for perpendicular magnetic fields and weakly disordered films, $\epsilon_{F}\tau\gg1$, where $\epsilon_{F}$
is the Fermi energy.\cite{Larkin04} For parallel magnetic fields
there is a minimum thickness $d_{clog}=\xi_{0}/\epsilon_{F}\tau$
below which paramagnetic effects start to dominate. This is known
as the Clogston limit,\cite{Clogston62,Chandrasekhar62} and we will assume that it is not reached, 
i.e., that $d\gg d_{clog}$. We note that recently 
the fluctuation conductivity in the opposite paramagnetic limit has also been addressed theoretically. \cite{Khodas12}

\subsubsection{Phase diagram}

\begin{figure}[tb]
\includegraphics[width=0.9\columnwidth]{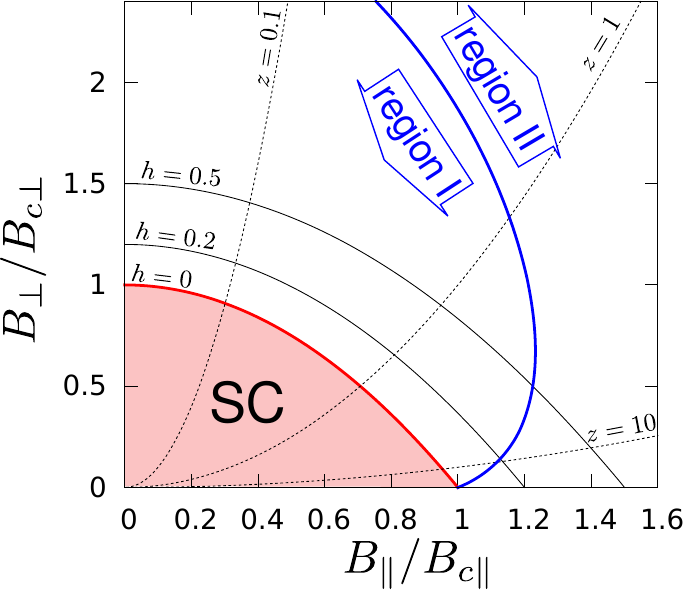}
\caption{(Color online) Phase diagram of the film in a tilted magnetic field at zero temperature. The diagram is parametrized by parameters $h$ and $z$, illustrated by lines of constant $h$ (solid) and constant $z$ (dashed). SC denotes the region of superconductivity. Region I and II refer to the asymptotic regions in which we evaluate the corrections to conductivity. They are separated by the line $h=2/(1+z)$. The sharp angular dependence of the negative correction to conductivity discussed below is a result of the pinching of region II near $B_{c\parallel}$.}
\label{fig:Phase-diagram} 
\end{figure}

The experimental phase diagram of the film is three-dimensional.
It is spanned by the temperature and the magnetic field, which is
further characterized by its magnitude $B$ and direction $\theta$.
A cut of the mean field phase diagram for zero temperature is displayed
in Fig.~\ref{fig:Phase-diagram}. The phase boundary of the superconducting
film is determined by the pair-breaking parameter $\alpha$,
which quantifies the effectiveness of the magnetic field to suppress
superconductivity.\cite{Maki69} In the considered case of a tilted
magnetic field, the pair-breaking parameter is the sum of the contributions due to the parallel and perpendicular field components,
$\alpha=\alpha_{\parallel}+\alpha_{\perp}$, where 
\begin{align}
\alpha_{\perp} & =DeB\sin\theta\label{eq:pair-breaking-sum-of-perp-par-1}\\
\alpha_{\parallel} & =\frac{Dd^{2}}{6}\left(eB\cos\theta\right)^{2}.\nonumber 
\end{align}
Here, $D=\frac{1}{3}v_{F}^{2}\tau$ is the electronic diffusion constant
of the material, where $v_{F}$ is the Fermi velocity.

The critical pair-breaking parameter $\alpha_{c}$, which separates
the normal phase with $\alpha>\alpha_{c}$ from the superconducting
phase with $\alpha<\alpha_{c}$, is temperature-dependent. It is implicitly
defined by the equation 
\begin{equation}
\ln\left(\frac{T}{T_{c0}}\right)-\psi\left(\frac{1}{2}\right)+\psi\left(\frac{1}{2}+\frac{\alpha_{c}\left(T\right)}{2\pi T}\right)=0,\label{eq:ag-formula}
\end{equation}
where $\psi\left(x\right)$ is the digamma function.\cite{Abramowitz72}
This equation has no solution for temperatures $T>T_{c0}$, for which
the system is a normal metal. By applying the asympotic expansion $\psi\left(x\right)\approx\ln x$
for large $x$, one finds that for zero temperature 
$\alpha_{c0}=\frac{\pi T_{c0}}{2\gamma}$, where $\ln \gamma \approx 0.577$ is the Euler constant.

Once the solution $\alpha_{c}\left(T\right)$ is known for arbitrary
temperatures, the phase boundary can be constructed. It is a two-dimensional
surface in the three-dimensional phase diagram spanned by the perpendicular
and parallel components of the magnetic field $B_{\perp}$ and $B_{\parallel}$
and by temperature. The phase boundary is determined by the equation
\be
\alpha_{\perp}+\alpha_{\parallel}=\alpha_{c}\left(T\right).\label{eq:alpha}
\ee

In order to find the critical field as a function of temperature
at a fixed angle $\theta$, for example, one should insert the expressions for $\alpha_\perp$ and $\alpha_\parallel$ of Eq.~\eqref{eq:pair-breaking-sum-of-perp-par-1} into Eq.~\eqref{eq:alpha} and obtains a quadratic equation for the
critical field strength $B_{c}(\theta,T)$. For a cut at constant temperature
$T\leq T_{c0}$, it is convenient to present Eq.~\eqref{eq:pair-breaking-sum-of-perp-par-1} in the form\cite{Harper68,Vedeneev02}
\begin{equation}
\left(\frac{B_{c}\left(\theta,T\right)\cos\theta}{B_{c\parallel}(T)}\right)^{2}+\frac{B_{c}\left(\theta,T\right)\sin\theta}{B_{c\perp}(T)}=1,\label{eq:BB}
\end{equation}
where $B_{c\perp}$ and $B_{c\parallel}$ are the (temperature-dependent)
critical fields for $\theta=\pi/2$ and $\theta=0$, respectively.
They can be obtained by setting $\alpha_{\parallel,\perp}=\alpha_{c}(T)$ and resolving for $B$. Eq.~\eqref{eq:BB} describes a parabolic phase boundary in the $(B_{\perp}-B_{\parallel})$-plane.

So far, the effect of a finite dephasing time was not included. As pointed out before, it may be accounted for by a shift in the eigenvalues $\alpha_n\rightarrow \alpha_n+1/2\tau_\phi$. Therefore, the condition for the mean-field transition can be written as $\alpha=\tilde{\alpha}_c(T)$, where $\tilde{\alpha}_c(T)=\alpha_c(T)-1/2\tau_\phi$ is modified due to the presence of dephasing effects. If the dephasing time is weakly magnetic-field dependent, then its main effect on Eq.~\eqref{eq:BB} is to renormalize the critical fields $B_{c\parallel}$ and $B_{c\perp}$.

We note that $B_{c\perp}(T=0)$ coincides with the nucleation critical
field $B_{c2}$. For low temperatures, $T\rightarrow0$, the two critical fields $B_{c\parallel}$ and $B_{c\perp}$ are related by 
\begin{equation}
\frac{B_{c\parallel}(T=0)}{B_{c\perp}(T=0)}=\frac{1}{d}\sqrt{\frac{6D}{\alpha_{c0}}}=4.16\times\frac{\xi_{0}}{d},
\end{equation}
where $\xi_{0}$ is the zero-temperature coherence length.\cite{Larkin04}

\subsubsection{Parameterization for the vicinity of the quantum critical line}
When formulating the results for the film in a tilted magnetic field below, we will specifically discuss the vicinity of the
quantum critical line in the phase diagram, i.e., we concentrate on
low temperatures, $t=T/T_{c0}\ll1$. For the fluctuation
conductivity, the regime of small temperatures is particularly interesting.
This regime displays the phenomenon of the non-monotonic magnetoresistance (NM),
as was first shown for the perpendicular magnetic field in Ref.~\onlinecite{Galitski01}
and for the parallel magnetic field in Ref.~\onlinecite{Lopatin05,Shah07}.
Here, we will discuss this phenomenon for magnetic fields tilted at
arbitrary angles $\theta$.

For a fixed temperature, the phase boundary has the shape of a parabola,
cf. Eq.~\eqref{eq:BB}. In order to present the results for the fluctuation
corrections, we introduce a parametrization of the region close to
this critical line, which will be described in the following. 

First, choose an arbitrary angle $\theta$ between 0 and 90 degrees.
Eq.~\eqref{eq:ag-formula} and \eqref{eq:BB}
determine the critical field strength $B_{c}\left(\theta\right)$,
at which, for a given angle, the system undergoes the phase transition.
The strength of the magnetic field can then be measured by the relative
distance $h$ to the phase boundary: 
\begin{equation}
h=\frac{B-B_{c}\left(\theta\right)}{B_{c}\left(\theta\right)},
\end{equation}
with $h=0$ corresponding to a point on the critical line and $h>0$
corresponding to a point in the normal region of the phase diagram.

In order to parametrize the angle $\theta$, it turns out to be useful to introduce another dimensionless number $\mymu$, which is
defined as the ratio between the two projected pair-breaking parameters 
\begin{equation}
\mymu\left(B,\theta\right)=\frac{\alpha_{\parallel}}{\alpha_{\perp}}=\frac{eBd^{2}}{6}\frac{\cos^{2}\theta}{\sin\theta}.\label{eq:mu-definition}
\end{equation}

As can be seen from the definition, $\mymu$ is directly related to
the angle, with $\mymu=0$ corresponding to perpendicular field and
$\mymu=\infty$ corresponding to parallel field. The parameters $h$
and $\mymu$ can be used instead of $B$ and $\theta$ in order to
define a point in the phase diagram. In fact, one can consider $h$
and $\mymu$ as a new curvilinear coordinate system of the phase space
that is aligned along the phase boundary of the system, as depicted
in Fig. \ref{fig:Phase-diagram}. 

Whenever the vicinity of the transition line in the $(B_{\perp}-B_{\parallel})$-plane
for a fixed temperature $T\ll T_{c0}$ is considered, the $B$ dependence
of $\mymu$ may be neglected and one may approximate $\mymu\approx\mymu\left(B=B_{c}\left(\theta,T\right),\theta\right)$.
This quantity can be determined experimentally (without explicit reference
to the thickness $d$):

\begin{equation}
\mymu\left(B=B_{c}\left(\theta,T\right),\theta\right)=\frac{B_{c\perp}(T)}{B_{c}\left(\theta,T\right)\sin\theta}-1.
\end{equation}
For almost parallel magnetic field, $\theta\ll1$, and for $T\rightarrow 0$,
one finds 
\begin{equation}
\mymu=\frac{B_{c \perp}(T=0)}{B_{c \parallel}(T=0)}\frac{1}{\theta}=0.24\times\frac{d}{\xi_{0}}\frac{1}{\theta}.
\label{eq:mu-for-small-angles}
\end{equation}

\subsubsection{Results for the fluctuation conductivity}

\label{sec:resforthefluctcond}

The kinetic equation approach employed in this paper leads to a rather natural classification of the three distinct contributions to fluctuation conductivity according to the underlying physical mechanisms. We distinguish the density of states contribution, $\delta \sigma_{dos}$, the anomalous Maki-Thompson term $\delta \sigma_{an}$, and the contribution of the fluctuating supercurrent $\delta \sigma_{sc}$, as discussed in the introduction. It is worth mentioning that this classification differs from the conventional diagrammatic scheme. For the parallel magnetic field case, the precise correspondence between the two formalisms is worked out in Sec.~\ref{sec:comparison}, and illustrated in Tab.~\ref{tab:table}.

We will now state results for the fluctuation conductivity of the film in the
vicinity of the quantum critical line, i.e., for $T\ll T_{c0}$ ($t\ll 1$). Let us stress
again that the general results displayed in Sec.~\ref{subsec:genresults} cover the entire
normal part of the phase diagram [with the exception of the region of
strong fluctuations very close to the transition]. Here, we focus on the
low-temperature regime since it displays the interesting phenomenon of the
NM. Whenever possible, we will discuss the origin of the different corrections according to the classification into density of states, anomalous Maki-Thompson and supercurrent contributions. 

When formulating our results we make use of the parameterization of the
phase diagram in terms of the parameters $h$, $\mymu$ and $T$ introduced above, see Fig.~\ref{fig:Phase-diagram}. For the low temperature regime, it can be expected that the presence of $\tau_\phi$ in the fluctuation propagator mainly leads to a shift in the critical line in the $(B_\perp-B_\parallel)$ plane. We assume that this shift has already been performed. At the same time, we neglect $\tau_\phi$ in the Cooperon, because at low $T$ and in the vicinity of the critical line, the Cooperon is not singular. This is why the presented formulas will not contain any explicit reference to $\tau_\phi$.

Comparatively simple semi-analytical expressions can be
found in two regimes. In region I, defined by the relation
$h\ll\frac{2}{1+\mymu}$, Landau level quantization of the Cooper pair
motion is crucial. Indeed, the dominant contribution to fluctuation
conductivity in this regime originates from fluctuations of the lowest Landau
level, since these fluctuations become singular at the transition. In regime
II, for which $h\gg\frac{2}{1+\mymu}$, the spectrum may be
approximated by a continuum for the purpose of the calculation. The reason is
that in this regime either, for small $z$, the distance to the transition line
is comparatively large and the fluctuations of all levels are non-singular or,
for large $z$, the magnetic field is almost parallel and the distance between
adjacent Landau levels becomes very small. 
The two regions are displayed in the diagram of Fig.~\ref{fig:Phase-diagram}.

With the only exception of very small angles $\theta$, the system is in region
I of the phase diagram when approaching the transition line, compare
Fig.~\ref{fig:Phase-diagram}. In turn, for large magnetic fields far from the transition
$h\gg 2$, the system is in region II independent of the
value of $\mymu$. The cross-over angle between the two regimes near criticality
can be estimated from Eq.~\eqref{eq:mu-for-small-angles}:  Region II is reached
only for very small angles 
\begin{equation}
        \theta \ll 0.12 \times \frac{d}{\xi_{0}} h. 
        \label{approximate-boundary-for-region-ii}
\end{equation}
As an illustration, for a film of thickness $d=0.3\,\xi_{0}$, fairly close to the transition $h=0.1$, the cross-over occurs at an angle $\theta$ of about $0.21^\circ{}$.

In the following, we will discuss the two regions separately, starting from
region I. After stating the results, we will provide a qualitative discussion of the behavior in the two regions.
\paragraph{Region I:} As a special example, region I contains the case of a strictly
perpendicular magnetic field, $\mymu=0$, in the vicinity of the transition.
This case has first been treated in Ref.~\onlinecite{Galitski01} (see also Ref.~\onlinecite{Tikhonov12}). Our results
for region I can be viewed as a generalization of these previous results to
non-perpendicular angles. 

The general formulas stated in Sec.~\ref{subsec:genresults}
involve an integration over an internal frequency and the summation over Landau
level indices. As was already noticed in Ref.~\onlinecite{Galitski01}, the most
singular contribution in the vicinity of the transition stems from the lowest
Landau levels (LL) only. Correspondingly, we consider the contributions due to
the singular LL and due to the higher Landau levels (HL) separately. It turns
out that for the HL a continuum approximation is sufficient. Furthermore, for
both LL and HL contributions we perform a separation into a thermal correction
(T), which vanishes for $T\rightarrow0$, and a quantum correction (0), which is
temperature-independent and thereby persists even in the limit $T\rightarrow
0$. As a result, the corrections to conductivity may be presented in the
following form:
\begin{align} \delta\sigma_{\text{I}} &
        =\delta\sigma_{0,LL}+\delta\sigma_{T,LL}+\delta\sigma_{0,HL}.
        \label{eq:contributions-to-region-i}
\end{align}
Here, the thermal contribution $\delta\sigma_{T,LL}$ reads as follows 
\begin{equation}
        \delta\sigma_{T,LL}= \frac{e^{2}}{\pi^{2}}
        \left( \alpha \tilde{I}_{\alpha} \left( r \right) +\beta I_{\beta} \left( r \right) \right),
        \label{eq:galitskii-larkin-general-alpha-beta}
\end{equation}
where
\begin{eqnarray}
        \tilde{I}_{\alpha}&=& \ln r-\frac{1}{2r}-\psi\left(r\right),\\
        I_{\beta}&=& r\psi'\left(r\right)-\frac{1}{2r} - 1,
        \label{eq:galitskii-larkin-ia-ib}
\end{eqnarray}
and we have abbreviated $r=\frac{h}{2\gamma t}$. This contribution is very similar in structure to the result for the perpendicular magnetic field case derived in Ref.~\onlinecite{Galitski01}. It differs mainly in two respects. First, we omitted the term $\ln h$ from $\tilde{I}_{ \alpha}$ as it does not vanish as $T\rightarrow 0$ and is therefore part of the quantum contribution $\delta \sigma_{0,LL}$ to be discussed below, cf. Eq.~\eqref{eq:dslllimit}. Second, the prefactors $\alpha$ and $\beta$ are now $\mymu$-dependent, i.e., they depend on the angle $\theta$,
\begin{equation}
\alpha=\frac{1}{3+\mymu}-\frac{1}{1+\mymu},\quad\beta=\frac{5+2\mymu}{3+\mymu}+\frac{1}{1+\mymu}.
\end{equation}
In the case of a perpendicular magnetic field, the coefficients reduce to the
previously derived $\alpha=-2/3$ and $\beta=8/3$.\cite{Galitski01,Tikhonov12}
Interestingly, however, in approaching the transition, $h \rightarrow 0$, we find that
$\delta\sigma_{\text{T,LL}}\approx \frac{2\gamma e^{2}}{\pi^2}\frac{t}{h}$,
meaning that the $\mymu$-dependence of
the general formula drops out in this limit.  When approaching the transition
at any finite temperature, $\delta\sigma_{T,LL}$ eventually becomes the
dominant contribution. It then resembles the well-known
Aslamazov-Larkin fluctuation correction. \cite{Aslamazov68, Galitski01} 

The quantum contribution due to the LL reads
\begin{align}
        \delta\sigma_{0,LL}&=\frac{e^{2}}{\pi^{2}}\left[
        \frac{1}{1+\mymu}\text{Li}\frac{1}{1+h} 
        + G\left( h,\frac{2\left( 1+h \right)}{\mymu + 1} \right) \right],\label{eq:sigma0LL}
\end{align}
where Li is the logarithmic integral function.\cite{Abramowitz72} The first term stems from the density of states correction $\delta \sigma_{dos}$, and the second part is due to the
supercurrent correction $\delta \sigma_{sc}$. The function $G\left( h,a \right)$ is defined by the integral
\begin{eqnarray}
        G(h,a)= \frac{1}{2}\int_{1+h}^{\infty}dx \;\frac{a}{x(x+a)}\left(\frac{1}{\ln x}-\frac{1}{\ln x+a}\right)
        \label{eq:definition-g-function}.
\end{eqnarray}

For $h\rightarrow 0$, the contribution $\delta\sigma_{0,LL}$ can be seen to
reduce to 
\be
\delta\sigma_{0,LL}\approx - \frac{e^{2}}{\pi^{2}} \alpha \ln \frac{1}{h} \qquad\quad (h\rightarrow 0),\label{eq:dslllimit}
\ee
which corresponds to the quantum term in the formula found by Galitski and Larkin.\cite{Galitski01}


The last contribution in Eq.~(\ref{eq:contributions-to-region-i}), $\delta \sigma _{0,HL}$, was omitted in Ref.~\onlinecite{Galitski01}, and
is obtained by considering the higher Landau levels. For this term, one may use the continuum approximation for the sum over Landau levels. This formally corresponds to the limit $\mymu \rightarrow 0$, i.e., this term is only very weakly $\mymu$-dependent. In addition, it is not singular when approaching the transition. Formally, the sum over higher Landau levels is very weakly (doubly logarithmically) divergent, so that it becomes necessary to introduce a high-energy cut-off $\Lambda$ and to take into account only modes with $\alpha_{n}<\Lambda$. As our theory is based on the diffusion approximation, the cut-off can be
chosen to be of the order of the transport scattering rate,
$\Lambda\simeq1/\tau$. 

The quantum correction from higher Landau levels can then be written as the sum of two integrals:
\be
	&&\delta\sigma^{(dos)}_{0,HL}=\\
	&&-\frac{e^2}{2\pi^{2}}\int_{0}^{\infty}dy\int_{0}^{K}dx\ 
	\frac{1}{\left(a+x+y\right)^{2}\ln\left(a+x+y\right)}, \no\\
	&&\delta\sigma^{(sc1)}_{0,HL}=\\
	&&\frac{e^2}{2\pi^{2}}\int_{0}^{\infty}dy\int_{0}^{K}dx\ 
	\frac{x}{\left(a+x+y\right)^{3} \ln^{2}\left(a+x+y\right)}.\no
	\label{eq:zero-temp-high-ll}
\ee
where we abbreviated $a=(1+h)^2$, and $K$ is a dimensionless cutoff given by $K=\frac{\Lambda}{2\alpha_{c0}}$. These integrals can be explicitely solved in terms of the logarithmic integral $\li$, see Eq.~\eqref{eq:zero-temp-high-ll-as-li} in Sec.~\ref{sec:tilted}.

As a final remark concerning the correction in region I, the thermal contribution originating from higher Landau levels, $\delta \sigma_{T,HL}$, has been omitted from formula (\ref{eq:contributions-to-region-i}) as it is regular and small.

\paragraph{Region II:} We now turn to the discussion of region II, where Landau levels are so close
that they can entirely be treated in the continuum approximation. In that case, the special significance of the
lowest Landau level is lost. As mentioned earlier, the limit of taking a continuous spectrum corresponds to the limit $\theta\rightarrow 0$.
The results for a film in a strictly parallel field have been found previously in the diagrammatic technique,\cite{Lopatin05,Shah07} and we agree with these results. For a more detailed discussion of the comparison, we refer to Sec.~\ref{sec:comparison}.

We separate the total correction to conductivity in region II into thermal and quantum contributions:
\begin{eqnarray}
        \delta\sigma_{\text{II}}&=& \delta\sigma_{T,HL}+\delta\sigma_{0,HL}.
        \label{eq:contributions-for-region-ii}
\end{eqnarray}
The expression for the quantum contribution $\delta\sigma_{0,HL}$ is the same in both regimes I and II and has already been stated above in Eq.~(\ref{eq:zero-temp-high-ll}). The thermal part is dominated only by the supercurrent correction $\delta \sigma_{sc}^{(2b)}$ in the classification introduced in Sec.~\ref{subsec:genresults} [in the conventional classification, it originates from the Aslamazov-Larkin diagram]. It can be written in the form \cite{Shah07}
\begin{eqnarray}
        \delta\sigma_{T,HL} &=& \frac{4e^2}{\pi}F\left( \eta \right), \quad  \eta =  \frac{\alpha-\alpha_{c}\left( T \right)}{T}.
        \label{eq:write-ds-in-terms-of-scaling-function}
\end{eqnarray}
For sufficiently small $h$, we can approximate $\eta=2\pi r$.  The function $F$ in the previous formula is defined by

\begin{equation}
	F(\eta) =\frac{\eta}{4\pi^{2}}\int_{\eta/\pi}^{\infty}\frac{\psi'\left(x\right)dx}{x}-\frac{1}{4\pi}-\frac{1}{16\eta}.\label{eq:F}
\end{equation}

It may be evaluated numerically, the asymptotic expansion for large and small values of the argument gives 
\begin{equation}
F\left(\eta\right)=\begin{cases}
\frac{\pi}{72\eta^{2}} & \mbox{for }\eta\gg1\\
\frac{1}{16\eta} & \mbox{for }\eta\ll1.
\end{cases}\label{eq:Flimits}
\end{equation}
We now turn to a more qualitative discussion. 

\subsubsection{Qualitative discussion: Film in a tilted magnetic field}

\paragraph{Region I:} The two dominant corrections in region I are $\delta\sigma_{T,LL}$ and $\delta\sigma_{0,LL}$. They are of different sign when
approaching the transition. For small $h\ll t$, in the so-called thermal
regime, $\delta\sigma_{T,LL}$ dominates. All the corrections $\delta \sigma_{dos}$, $\delta \sigma_{an}$ and $\delta\sigma_{sc}$ contribute to $\delta \sigma_{T,LL}$ and the net result is a positive correction to conductivity as naively expected when approaching the superconducting state. However, the presence of the term $\delta\sigma_{0,LL}$, which is negative and dominates in the so-called quantum regime $h\gg t$,
leads to a local minimum in $\delta\sigma$ as a function of $h$ at the
cross-over between the thermal and the quantum regime. The result is the NM.\cite{Galitski01,Baturina05} The physical origin is the negative density of states correction, which competes with the positive supercurrent correction in the quantum regime $h\gg t$, but is numerically larger. The anomalous Maki-Thompson term is ineffective in this regime. The physical mechanism underlying the phenomenon of the NM in region I can therefore be expressed in simple terms. Cooper pairs form, but are comparatively immobile as a consequence of their quantized spectrum (Landau level quantization). The decrease in conductivity due to a reduction of the density of states of quasiparticles may therefore overcome the increase in conductivity caused by the fluctuation supercurrent carried by these Cooper pairs.

\begin{figure}
\includegraphics[width=\columnwidth]{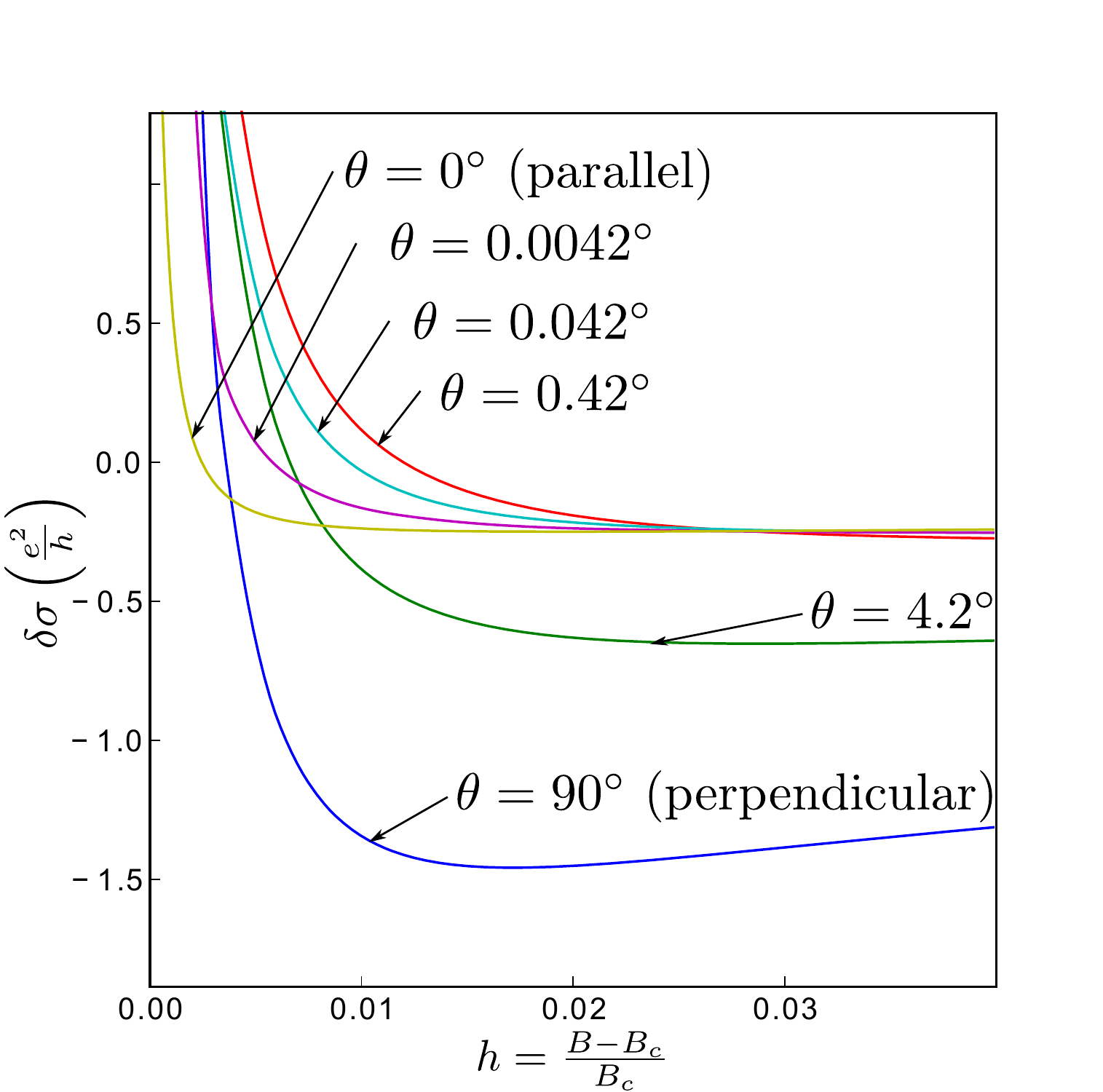}
\caption{(Color online) The correction to conductivity of a disordered film at very low temperatures ($T=0.005\,T_{c0}$, $d/\xi_0=0.3$).
The transition is approached by tuning the magnitude of the magnetic field at a fixed angle $\theta$.
When one tilts the angle starting from the perpendicular magnetic field case, the dominant effect is a reduction of the negative correction to conductivity. When tilting further towards parallel fields a second effect sets in, namely a sudden decrease of the thermal contribution. This effect becomes visible mainly for small $h$.}
\label{fig:numerical-results}
\end{figure}

\paragraph{Crossover between regions I and II:} The importance of the term $\delta\sigma_{0,HL}$ lies in the fact that,
irrespective of the precise choice of the cut-off, it is negative and weakly angular dependent. Deep in region
I, the phenomenon of the NM is largely determined by the interplay of
$\delta\sigma_{0,LL}$ and $\delta\sigma_{T,LL}$ as discussed above, and $\delta
\sigma_{0,HL}$ is of minor importance. As one approaches the cross-over regime
between regions I and II at small angles, however, the negative correction due
to $\delta \sigma_{0,LL}$ diminishes, as it is proportional to $\alpha$, and
$\delta \sigma_{0,HL}$ becomes more relevant. As will be discussed in more
detail below, in region II the NM still exists only due to the existence of
the negative contribution $\delta \sigma_{0,HL}$. For the accurate description
of the cross-over regime itself, a numerical evaluation is necessary and
results are displayed in Fig.~\ref{fig:numerical-results}. Let us remind at this point that the negative 
correction $\delta \sigma_{0,HL}$ stems from the density of states and
 supercurrent corrections $\delta \sigma_{dos}$ and $\delta \sigma_{sc}$. Again, 
 the negative density of states contribution dominates.
 
 Another interesting observation can be made in the thermal regime $h\ll t$, 
which can be reached in both regions by approaching the transition. It is
characteristic for this regime that the divergent thermal contributions,
$\delta\sigma_{T,LL}$ and $\delta\sigma_{T,HL}$, dominate. As mentioned before,
very close to the transition, the system is in region I for almost all angles.
The cross-over to region II only happens very close to $\theta=0$. 
It is thus interesting to note that for $h\ll t$, the asymptotic expansion of
$\delta\sigma_{T,LL}$ contributing to $\delta\sigma_\text{I}$
is $\frac{2\gamma e^{2}}{\pi^{2}}\frac{t}{h}$, while
the asymptotic expansion of $\delta\sigma_{T,HL}$, which contributes to $\delta
\sigma_\text{II}$, is $\frac{\gamma e^{2}}{4\pi^{2}}\frac{t}{h}$. Thus, while
crossing over from region I to region II in the thermal regime, the fluctuation
correction
drops to about $1/8$ of its value for small $h$.

\paragraph{Region II:} As mentioned above, $\delta\sigma_{0,HL}$ is regular at the transition, and
slowly decreases when moving towards the normal regime. On the other hand,
the thermal contribution $\delta\sigma_{T,HL}$ diverges when approaching the
transition. As demonstrated in Ref.~\onlinecite{Lopatin05,Shah07}, the interplay of these two
contributions also results in a NM. From the preceding discussion it is clear that the non-monotonic behavior of the magnetoresistance has a different origin for parallel and perpendicular magnetic fields. For parallel fields, the negative correction comes from $\delta\sigma_{0,HL}$. For perpendicular fields, it originates from $\delta\sigma_{0,LL}$. In both cases, however, these negative corrections stem from a competition of density of states and supercurrent terms, for which the negative $\delta \sigma_{dos}$ dominates over the positive $\delta \sigma_{sc}$. 

This concludes the discussion of the main results. In the following sections, details of the derivation will be presented.

\section{General formalism}
\label{sec:general}
\global\long\def\chempot{\mu}

In this section, we describe the formalism underlying the results presented in Sec.~\ref{subsec:genresults}. We derive a quasiclassical kinetic equation, the so-called Usadel equation. The form of this equation is slightly nonstandard in order to allow for the inclusion of fluctuations outside the superconducting regime. The formalism used here was introduced in Ref.~\onlinecite{Tikhonov12} for a film in a perpendicular magnetic field. We generalize it here in order to be able to treat parallel field components as well. The main difference to Ref.~\onlinecite{Tikhonov12} is that a more general set of eigenfunctions and eigenvalues is considered. Otherwise, the derivation is analogous to the one presented in Ref.~\onlinecite{Tikhonov12}. We describe the main steps here in order to coin the notation and to make the paper self-contained. Previous related works include Ref.~\onlinecite{Volkov98}, in which the Usadel equation was used for the calculation of fluctuation corrections close to $T_{c0}$ 
and Refs.~\onlinecite{Levchenko07} and \onlinecite{Petkovic13}, where the Keldysh nonlinear sigma-model was employed for the calculation and again, only the vicinity of $T_{c0}$ was studied. The latter two works are based on the Keldysh sigma-model approach for superconductors presented in Ref.~\onlinecite{Feigelman00}; the connection between the different formalisms is that the Usadel equation is the saddle point equation of the sigma-model. 

\subsection{Microscopic model}
We start from the Keldysh action for electrons with BCS short-range interaction. The interaction in the Cooper channel is already decoupled by means
of a Hubbard-Stratonovich transformation, leading to the action
\begin{eqnarray}
S[\chi,\Delta]=\int d{\bf r}\int_{\mathcal{C}}dt\ \sum_{\alpha=\uparrow,\downarrow}\chi_{\alpha}^{*}\left(i\partial_{t}-{h}+\ce{\chempot}\right)\chi_{\alpha}\no\\
+\Delta^{*}\chi_{\downarrow}\chi_{\uparrow}+\Delta\chi_{\uparrow}^{*}\chi_{\downarrow}^{*}+\frac{\nu}{\lambda}\Delta^{*}\Delta.
\end{eqnarray}
Here, $\chi_{\uparrow}(x)$ and $\chi_{\downarrow}(x)$ are Grassmann fields
describing the spin-up and spin-down components of the electrons at the space-time point $x=(\bfr,t)$,
${h}$ is the single-particle Hamiltonian, $\Delta(x)$ is the superconducting
order parameter field, $\nu$ is the density of states of electrons
per spin direction at the Fermi surface, and $\lambda>0$ is a dimensionless
parameter, which determines the strength of electron-electron attraction
in the Cooper channel. The integral in time is along the closed Keldysh
contour $\mathcal{C}$.\cite{Kamenev11} The single particle Hamiltonian 
\begin{equation}
{h}=-\frac{1}{2m}\left(\boldsymbol{\nabla}-ie{\bf A}\left({\bf r}\right)\right)^{2}+U\left({\bf r}\right)+e\varphi\left({\bf r}\right)
\end{equation}
 includes coupling to a vector potential ${\bf A}$ and an
external scalar potential $\varphi$, as well as an impurity potential
$U$. In this work, we assume a constant
potential gradient $\varphi\left({\bf r}\right)=-{\bf E}\cdot{\bf r}$,
where ${\bf E}$ is the electric field. 

We separate fields on the forward and backward branch
of the Keldysh contour. We further introduce the Nambu (N) particle-hole space, and perform the Keldysh rotation. This means that
we describe our system by the four-component fields $ \Psi =(\psi_{1},\psi_{2})^T_K,\ \psi_{i}=(\chi_{\uparrow i},\chi_{\downarrow i}^{*})^T_N$ and $\Psi^{\dagger}  =(\psi^\dagger_{1},\psi^\dagger_{2})_K,\ \psi_{i}^{\dagger}=(\chi^*_{\uparrow i},-\chi_{\downarrow i})_N$, where K labels the Keldysh space.
This transformation turns the action into
\begin{eqnarray}
S\left[\Psi,\check{\Delta}\right]&=&\int dx\,\Psi^{\dagger}(i\hat{\tau}_{3}\partial_{t}-\check{H}+\chempot+\check{\Delta})\Psi\no\\
&&\qquad-\frac{2\nu}{\lambda}\mbox{tr}(\check{\Delta}^{\dagger}\sigma_1{\check{\Delta}}).\label{eq:suco_action}
\end{eqnarray}
Here and in the following, $\hat{\tau}_{i}$ denote Pauli matrices in particle-hole
space, while $\hat{\sigma}_{i}$ denote Pauli matrices in Keldysh space. 
The trace is taken in the four-dimensional product space. Matrices in this product space are denoted as $\check{\Delta}$, $\check{H}$, etc.. The time-integration covers the real axis. The single-particle
Hamiltontian $\check{H}$ is obtain from the Hamiltonian $h$
by replacing ${\bf A}$ by ${\bf A}\hat{\tau}_{3}$. The $4\times 4$ matrix $\check{\Delta}$ contains the classical (c) and quantum (q) components of the order parameter field $\Delta$ as $\check{\Delta}=\hat{\Delta}_{0}\hat{\sigma}_{0}+\hat{\Delta}_{1}\hat{\sigma}_{1}$, where $\hat{\Delta}_i=\Delta_i\hat{\tau}_+-\Delta^*_i\hat{\tau}_-$, $\hat{\tau}_{\pm}=(\hat{\tau}_x\pm i\hat{\tau}_y)/2$ and $\vec{\Delta}=(\Delta_{c},\Delta_q)^T$.
\cite{Tikhonov12,Kamenev11}

In order to perform the quasiclassical approximation, it is useful to define the
electronic Green's function
\begin{equation}
i\check{G}_{\Delta}\left(x,x'\right)=\frac{\int\left[d\Psi\right]\Psi\left(x\right)\Psi^{\dagger}\left(x'\right)e^{iS\left[\Psi,\check{\Delta}\right]}}{\int\left[d\Psi\right]e^{iS\left[\Psi,\check{\Delta}\right]}},
\end{equation}
where the average is performed over the electronic degrees of
freedom only, keeping the field $\check{\Delta}$ fixed. In order to recover the
full Green's function $\check{G}$ one needs to average as
$\check{G}=\int\left[d\check{\Delta}\right]\check{G}_{\Delta}e^{iS_{GL}\left[\check{\Delta}\right]}$ with respect to the Ginzburg-Landau action $S_{GL}\left[\check{\Delta}\right]=-i\ln\int\left[d\Psi\right]e^{iS\left[\Psi,\check{\Delta}\right]}$.
In Keldysh space, the matrix Green's function has the typical triangular form
\begin{eqnarray}
\check{G}=\left(\ba{cc} G^R&G^K\\0&G^A\ea\right)\label{eq:gf-keldysh-structure}.
\end{eqnarray}
It is important to note that in general $G_{\Delta}$ does not have the same structure. If $\Delta_q\ne 0$, the element in the lower left corner, sometimes referred to as $G_{\Delta}^Z$, is not equal to zero. For the linear response calculation of the fluctuation conductivity in the Gaussian approximation, however, it turns out to be sufficient to put $G^Z_{\Delta}=0$ as it is proportional to higher powers of the order parameter field. This has been shown in Ref.~\onlinecite{Tikhonov12}. We will therefore work with the triangular Keldysh structure for $G_{\Delta}$.

In the presence of impurities, the Green's function $\check{G}$ needs to be
averaged over an ensemble of disorder configurations. Physical quantities can
be calculated with the help of the disorder averaged Green's function
$\left\langle \check{G}_{\Delta}\right\rangle_{dis}$. For films with a
dimensionless conductance $g\gg 1$, which we consider, it is legitimate to
average as $\left\langle
\check{G}\right\rangle_{dis}=\int\left[d\check{\Delta}\right]\left\langle\check{G}_{\Delta}\right\rangle_{dis}e^{i\left\langle
	S_{GL}\left[\check{\Delta}\right]\right\rangle_{dis}}$, i.e.,
separately for $G_{\Delta}$ and $S_{GL}$. 
Corrections originating from cross-correlations between the two terms would be
smaller by a factor $1/g$ than the quantum corrections that are the subject of
this paper.
The quasiclassical approximation may now be introduced for the disorder
averaged Green's function $\left\langle G_{\Delta}\right\rangle_{dis}$ at a
fixed order parameter configuration. 

\subsection{Quasiclassical approximation and the Usadel equation}

We assume that $\epsilon_{F}$ is the largest energy scale in the problem, meaning that
superconductivity, scattering and external fields only affect the
system close to the Fermi surface: $\Delta,\tau^{-1},\omega_{ext}\ll \epsilon_{F}$.
Thus, one may use the so-called quasiclassical approximation.\cite{Kopnin01} To this end one transforms
to Wigner coordinates and defines the quasiclassical Green's function by integrating over the distance to the Fermi surface 
\begin{equation}
\check{g}_{{\bf n}}\left({\bf r},t,t'\right)=\frac{i}{\pi}\int d\xi_{{\bf p}}\, \left\langle \check{G}_{\Delta}\right\rangle_{dis} \left({\bf r},{\bf p},t,t'\right).
\end{equation}
Here, ${\bf n}$ is a unit vector pointing in the direction
of ${\bf p}$, and $\xi_{\bf p}=\frac{p^{2}}{2m}-\chempot$.
The quasiclassical Green's function $\check{g}_{\bf n}$ obeys the Eilenberger equation.\cite{Eilenberger68}

In the limit of dirty superconducitivity, $T_{c0}\tau\ll1$, frequent scattering washes out the angular dependence of the Green's function, which allows to work with the simpler function $\check{g}=\int d{\bf n} \, \check{g}_{\bf n}$. The appropriate equation in this limit is the Usadel equation \cite{Usadel70,Kopnin01}
\begin{equation}
D{\hat{\partial}_{\bf A}}\left(\check{g}{\hat{\partial}_{\bf A}}\check{g}\right)-\left\{ \hat{\tau}_{3}\partial_{t},\check{g}\right\} +i\left[\check{\Delta}-\check{\varphi},\check{g}\right]=0.\label{eq:usadel-equation}
\end{equation}
Here, $D=v_{F}^{2}\tau/3$ is the electronic diffusion coefficient for diffusion in three dimensions and $\check{\varphi}$ is the external potential, which is considered as a matrix in Nambu and Keldysh space. 
We defined the derivative $\hat{\partial}_{\bf A}\check{g}=\nabla\check{g}-ie\left[{\bf A}\hat{\tau}_{3},\check{g}\right]$. Also note that one should read the appearing anti-commutator in an operator sense, $\left\{ \hat{\tau}_{3}\partial_{t},\check{g}\right\}(t_1,t_2)=\tau_3\partial_{t_1}g(t_1,t_2)-\partial_{t_2}g(t_1,t_2)\tau_3$. In this equation and the following, we consider $\check{g}$, $\check{\Delta}$ and $\check{\varphi}$
as matrices in Keldysh and Nambu space and in time. Thus a multiplication implies the matrix product of the $4\times 4$ matrix
as well as a convolution in time. The Usadel equation has to be supplemented
with the constraint $\check{g}^{2}=\check{1}$. The current in the sample can be expressed through $\check{g}$
as \cite{Rammer86}
\begin{equation}
{\bf j}=\frac{e\pi\nu D}{2}\mbox{tr}\left(\hat{\tau}_{3}\hat{\sigma}_{1}\left(\check{g}\boldsymbol{\hat{\nabla}}\check{g}\right)\right).\label{eq:current-from-quasiclassical}
\end{equation}

This procedure is an extension of the usual quasiclassical formalism
in the Keldysh technique; here, the formalism is applied without employing
the mean-field approximation for the field $\Delta$. This allows
us to treat the electronic system in the quasiclassical approximation
while fluctuations of the order parameter field $\Delta$ can still
be taken into account. 

In the rest of this section, we consider the normal side of the transition,
where fluctuations of the order parameter field are taken into account in the Gaussian approximation. We therefore linearize the Usadel equation using the solution $\Delta=0$ as a starting point. This allows to obtain an expression for the current as a functional
of the field $\Delta$, which can then be average using the correlation
function for $\Delta$. It is possible to obtain the corresponding fluctuation propagator from
the quasiclassical Green's function itself.

In analyzing fluctuations around the normal state of the metal, we
parametrize the Green's function as in Eq. \eqref{eq:gf-keldysh-structure},
using 
\begin{align}
g^{R} & =\left(\begin{array}{cc}
1-f\fc/2 & f\\
\fc & -1+\fc f/2
\end{array}\right),\label{eq:gr_parametrization}\\
g^{A} & =\left(\begin{array}{cc}
-1+\fb\fbc/2 & -\fb\\
-\fbc & 1-\fbc\fb/2
\end{array}\right),\label{eq:ga_parametrization}
\end{align}
while for the Keldysh component we set $g^{K}=g^{R}h-hg^{A}$. In these equations $g^K$, $g^R$, $g^A$, $h$, and $f$ are all functions of one spatial and two time coordinates, $g^K=g^K(\bfr,t,t')$ etc., and the product implies a convolution in time. The parametrization is consistent with the non-linear constraint $\check{g}^2=\check{1}$ up to third order in $f$. The functions
$f$,$\fc$, $\fb$, $\fbc$ are considered to be \emph{independent}
from each other; in fact, it turns out that $\fc$ is the conjugate
of $f$ only if $\Delta_{q}=0$. Inserting \eqref{eq:gr_parametrization}
and \eqref{eq:ga_parametrization} into the Usadel equation \eqref{eq:usadel-equation},
we obtain a linear equation for $f$, 
\begin{equation}
\mathcal{C}^{-1}f=V,
\end{equation}
where
\begin{eqnarray}
V_{t_1,t_2}(\bfr)=2i[\Delta_{c}(\bfr,t_1)\delta_{t_1-t_2}+h_{e}(\bfr,t_1-t_2)\Delta_{q}(\bfr,t_2)],\no\\
\end{eqnarray}
the operator $\mathcal{C}^{-1}$ is defined as 
\begin{eqnarray}
\mathcal{C}^{-1}&=&D\partial_{\bf A}^2-\left(\partial_{t_1}-\partial_{t_2}\right),\quad \partial_{\bf A}={\nabla}- 2ie{\bf A}(\bfr),\quad 
\end{eqnarray}
and $h_{e}$ will be introduced in Eq.~(\ref{eq:distr_function}) below. 
For $\fc,\fb,\fbc$ similar equations are obtained, which show that $\fb_{t_1,t_2}=-f_{t_2,t_1}$, $\fc_{t_1,t_2}=-\fbc_{t_2,t_1}$, while in general $\fc\neq\left(f\right)^{*}$. 

From the kinetic equation for $h$, to lowest order in $\Delta$, one
finds that the only deviation from the equilibrium distribution function
is a shift due to the local potential\cite{Tikhonov12}
\begin{eqnarray}
        h\left({\bf r},\epsilon\right)&=&\mbox{diag}\left(h_{e}(\bfr,\epsilon),h_{h}(\bfr,\epsilon)\right), \label{eq:distr_function}\\
        h_{e/h}(\bfr,\epsilon)&=&\mathcal{H}(\epsilon\mp  e\varphi\left({\bf r}\right)). 
\end{eqnarray}
Here, $\mathcal{H}(\epsilon)=\tanh(\epsilon/2T)$ is the fermionic equilibrium distribution function.

In order to solve the linearized Usadel equation for $f$, we introduce the Fourier transform 
\begin{eqnarray}
f_{t_1,t_2}=\int_{\epsilon_1\epsilon_2} f_{\epsilon_1\epsilon_2}\mbox{e}^{-i\epsilon_1 t_1+i\epsilon_2 t_2},
\end{eqnarray}
where $\int_{\epsilon}=\int d\epsilon/2\pi$. The same convention is used for the independent variables $f^*$, $\overline{f}$, $\fbc$, and for $V$. In a second step, we expand $f=\sum_{n}f_n\phi_n$, $f^*=\sum_n f^*_n\phi^*_n$, etc., and use that the spatial part of the operator $\mathcal{C}^{-1}$ is diagonal in the basis of eigenfunctions $\phi_n$ of Eq.~\eqref{eq:eigenvalue-equation-1}. It gives
\begin{eqnarray}
f_{n;\epsilon_1,\epsilon_2}=C_{n;\epsilon_1+\epsilon_2}\int d{\bf r}\; \phi_n^*(\bfr)\;V_{\epsilon_1,\epsilon_2}(\bfr),\label{eq:linearized-usadel-equation}
\end{eqnarray}
where we introduced the so-called Cooperon $C_n$
\begin{eqnarray}
C^{-1}_{n;\epsilon}=i\epsilon-2\alpha_n-1/\tau_\phi,\label{eq:cooperon}
\end{eqnarray}
including a phenomenological dephasing rate as discussed in Sec.~\ref{subsec:genresults}.

We now turn to the fluctuation propagator $\mathcal{L}$, which is directly related to the correlation
function of the field $\Delta$
\begin{align}
\left\langle \Delta_i(x_1)\Delta^*_j(x_2)\right\rangle & =\frac{i}{2\nu}\mathcal{L}_{ij}(x_1,x_2).
\label{eq:lij-definition}
\end{align}
The indices $i,j$ label the components of the vector $\vec{\Delta}$ defined in Eq.~(\ref{eq:suco_action}).
Conveniently, the fluctuation propagator can be found from
the quasiclassical Green's function.\cite{Tikhonov12} We restrict ourselves to the Gaussian approximation for the fluctuations of the order parameter field, i.e., we find the correlator by approximating $S_{GL}$ by its expansion to second order in $\Delta$. The resulting term contains the Greens function $\check{G}_{\Delta}$ evaluated at coinciding space points, and can thus be expressed through the quasiclassical Green's function $\check{g}$. One finds
\begin{multline}
\left(\mathcal{L}^{-1}\right)_{ij}\left({\bf r},{\bf r'},t,t'\right)=\pi\nu\frac{\delta}{\delta\Delta_{j}^{*}\left({\bf r'},t'\right)}\mbox{tr}\hat{\sigma}_{i}\hat{\tau}_{-}\check{g}\left({\bf r},t,t\right)\\
-\left(\frac{2\nu}{\lambda}\hat{\sigma}_{1}\right)_{ij}\delta\left({\bf r}-{\bf r'}\right)\delta\left(t-t'\right),\label{eq:finding the propagator}
\end{multline}
For the first term on the right hand side the identification $\sigma_{c}=1$ and $\sigma_{q}=\sigma_{1}$ is used. The fluctuation
propagator can thus be found from the solution of the Usadel equation. It is convenient to work in the basis of eigenfunctions $\psi_n$ of \eqref{eq:eigenvalue-equation-1} and to present the fluctuation propagator in the form
$\mathcal{L}(\bfr,\bfr')=\sum_{nm}\psi_n(\bfr)\mathcal{L}_{nm}\psi^*_m(\bfr')$.
To first order in the electric field, one finds the following matrix elements
\begin{align}
2\nu\mathcal{L}_{nm}^{R}\left(\omega\right) & =-i\left\langle \Delta_{c,n}\left(\omega\right)\Delta_{q,m}^{*}\left(\omega\right)\right\rangle \nonumber \\
 & =L_{n}\delta_{nm}-2{\bf E}{\bf r}_{nm}\e_{m}^{\prime}L_{n}L_{m},\\
2\nu\mathcal{L}_{nm}^{A}\left(\omega\right) & =-i\left\langle \Delta_{q,n}\left(\omega\right)\Delta_{c,m}^{*}\left(\omega\right)\right\rangle \nonumber \\
 & =L_{n}^{*}\delta_{nm}-2{\bf E}{\bf r}_{nm}\e_{n}^{*\prime}L_{n}^{*}L_{m}^{*},
\end{align}
and 
\begin{align}
2\nu\mathcal{L}_{nm}^{K}\left(\omega\right) & =-i\left\langle \Delta_{c,n}^{*}\left(\omega\right)\Delta_{c,m}\left(\omega\right)\right\rangle \nonumber \\
 & =\mathcal{B}\left(L_{n}-L_{n}^{*}\right)\delta_{nm}\nonumber \\
 & \quad-\mathcal{B}'{\bf E}{\bf r}_{nm}\left(L_{m}^{*}-L_{n}+\e_{m}L_{n}L_{m}^{*}-\e_{n}^{*}L_{n}L_{m}^{*}\right)\nonumber \\
 & \quad-2\mathcal{B}{\bf E}{\bf r}_{nm}\left(\e_{m}'L_{n}L_{m}-\e_{n}^{*\prime}L_{n}^{*}L_{m}^{*}\right).
\end{align}
Furthermore, ${\bf r}_{nm}$ are the matrix elements of the position operator given in Eq.~\eqref{eq:rnm}
and $\mathcal{B}\left(\omega\right)=\coth\left(\omega/2T\right)$ is the bosonic
equilibrium distribution function. $L_n$ is defined in Eqs.~\eqref{eq:L} and \eqref{eq:E}. In Eq.~\eqref{eq:E}, $\alpha_n$ should be replaced by $\alpha_n+1/2\tau_\phi$ in order to account for the finite dephasing rate introduced in Eq.~\eqref{eq:cooperon}.

In the next step, we insert the parametrization (\ref{eq:gr_parametrization}), (\ref{eq:ga_parametrization}) into formula (\ref{eq:current-from-quasiclassical}). From the normal-metal solution, for which $\hat{g}^R=-\hat{g}^A=\hat{\tau}_3$, one immediately finds the Drude expression for the current ${\bf j}^{(n)}=2\nu e^2 D{\bf E}$. Taking into account fluctuations up to second order in $f$, one obtains several terms with distinct structure. They can be grouped as follows\cite{Tikhonov12}
\begin{eqnarray}
{\bf j}^{(\text{dos})} &=&-\frac{e^{2}\pi \nu D{\bf E}}{2}\int_{\epsilon\epsilon'}\, \mathcal{H}'(\epsilon)\left(\fc_{\epsilon\epsilon'}f_{\epsilon'\epsilon}+f_{\epsilon\epsilon'}\fc_{\epsilon'\epsilon}\right),\quad\label{eq:jdos}\\
{\bf j}^{(\text{an})} &=&-{e^{2}\pi \nu D{\bf E}}\int_{\epsilon\epsilon'} \, \mathcal{H}'\left(\epsilon\right)\left(\fb_{\epsilon\epsilon'}\fc_{\epsilon'\epsilon}\right),\label{eq:jan}\\
{\bf j}^{(\text{sc})} &=&{e\pi \nu D}\int_{\epsilon\epsilon'}\, 
\mathcal{H}\left(\epsilon\right)\bigl(
f_{\epsilon\epsilon'}\nm\fc_{\epsilon'\epsilon}\no
+\nm\fc_{\epsilon\epsilon'}f_{\epsilon'\epsilon}\\
&&-\fc_{\epsilon\epsilon'}\np f_{\epsilon'\epsilon}
-\np f_{\epsilon\epsilon'}\fc_{\epsilon'\epsilon}\bigr).\label{eq:jsc}
\end{eqnarray}
where we abbreviated $\nabla_\pm=\nabla\mp2ie\mathbf{A}$. In the result \eqref{eq:jsc}, we dropped terms that contain $\mathcal{H}'(\epsilon)$. These terms vanish after averaging over fluctuations. 
The naming of these three components refers to density of states (dos), anomalous Maki-Thompson (an) and super-current (sc), respectively. 

Eqs.~(\ref{eq:jdos}), (\ref{eq:jan}) and (\ref{eq:jsc}) still need to be averaged over the fluctuating order parameter field.
To this end, one needs to express $f$, $f^*$, etc. in terms of $\Delta$ and
$\Delta^*$ and the Cooperon, as exemplified
Eq.~(\ref{eq:linearized-usadel-equation}). The averaging is then performed with
the help of Eq.~\eqref{eq:lij-definition}. The integral over the frequency
$\epsilon$ can be performed analytically. This step is straight-forward in
principle, but tedious, this is why we display some intermediate steps in App.~\ref{appendix_integration}. The final results for the fluctuation corrections are displayed in
Eqs.~\eqref{eq:dosansc} to \eqref{eq:general-al2b-1} in Sec.~\ref{subsec:genresults}.

\section{Comparison to the diagrammatic technique}
\label{sec:comparison}

\begin{table*}
\begin{tabular}{|c|c|c|c|}
\hline 
Class of diagrams    \bigstrut                                                      & terms in Ref.~\onlinecite{Shah07}, cf. Eq.~\eqref{eq:shah-lopatin-result}             & after integ. by parts                                                                                     & terms in Usadel technique, cf. Eq.~\eqref{eq:usadel-result-for-shah-lopatin}                        \\
\hline 

\includegraphics[width=2cm]{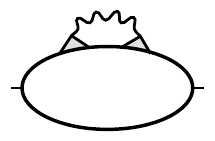}\bigstrut                    & \multirow{1}[4]{*}[0.5cm]{$\delta\sigma_{DOS}^{\left(sh,2c\right)}-A$ }\bigstrut   & \multirow{1}[2]{*}[0.5cm]{$\delta\sigma_{DOS}^{\left(sh,2c\right)}-A$}\bigstrut                          & \multirow{2}[4]{*}[0.25cm]{$\delta\sigma_{dos}$}\bigstrut \\

\cline{1-3} 

\multirow{2}[4]{*}{\includegraphics[width=2cm]{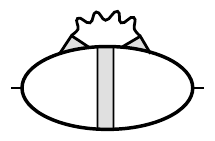}}         & \multirow{2}[4]{*}[-0.2cm]{$\delta\sigma_{DOS}^{\left(sh,3c\right)}+\frac{3}{2}B$}\bigstrut & \multirow{1}[2]{*}[-0.2cm]{$\frac{1}{2}A-\frac{1}{2}\delta\sigma_{DOS}^{\left(sh,2c\right)}$}\bigstrut\bigstrut   &                                                 \\[0.5cm]

\cline{3-4} 

                                                                           &                                                                                     & \multirow{1}[2]{*}{$-\delta\sigma_{sc}^{\left(1\right)}-\delta\sigma_{sc}^{\left(2a\right)}$}\bigstrut    & \multirow{4}[8]{*}{$\delta\sigma_{sc}$}\bigstrut\\

\cline{1-3} 

\multirow{2}[4]{*}{\includegraphics[width=2cm]{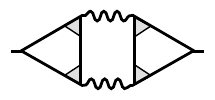}}\bigstrut   & \multirow{1}[2]{*}{$\delta\sigma_{AL}^{\left(cth\right)}$}\bigstrut                 & \multirow{1}[2]{*}{$4\delta\sigma_{sc}^{\left(1\right)}$}\bigstrut                                        &                                                 \\

\cline{2-3} 

                                                                           & \multirow{1}[2]{*}{$\delta\sigma_{AL}^{\left(sh\right)}$}\bigstrut                  & \multirow{1}[2]{*}{$2\delta\sigma_{sc}^{\left(2a\right)}+\delta\sigma_{sc}^{\left(2b\right)}$}\bigstrut   &                                                 \\

\cline{1-3} 

\multirow{2}[4]{*}{\includegraphics[width=2cm]{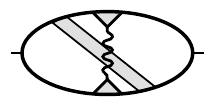}}\bigstrut & \multirow{2}[4]{*}{$3B$}\bigstrut                                                   & \multirow{1}[2]{*}{$-2\delta\sigma_{sc}^{\left(1\right)}$}\bigstrut                                       &                                                 \\

\cline{3-4} 

                                                                           &                                                                                     & \multirow{1}[2]{*}{$A$}\bigstrut                                                                          & \multirow{2}[4]{*}{$=0$}\bigstrut               \\

\cline{1-3} 

\multirow{2}[4]{*}{\includegraphics[width=2cm]{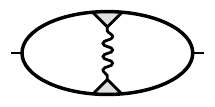}}\bigstrut & \multirow{1}[2]{*}{$-A$}\bigstrut                                                   & \multirow{1}[2]{*}{$-A$}\bigstrut                                                                         &                                                 \\

\cline{2-4} 

                                                                           & \multirow{1}[2]{*}{$\delta\sigma_{MT}^{\left(sh\right)}$}\bigstrut                  & \multirow{1}[2]{*}{$\delta\sigma_{MT}^{\left(sh\right)}$}\bigstrut                                        & \multirow{1}[2]{*}{$\delta\sigma_{an}$}\bigstrut \\

\hline
\end{tabular}
\caption{A comparison of the expressions for the fluctuation conductivity in a parallel magnetic field obtained from a diagrammatic calculation in Ref.~\onlinecite{Shah07} and from the Usadel equation approach in this manuscript. The LHS shows the classes of diagrams evaluated in Ref.~\onlinecite{Shah07} together with the corresponding terms given in \eqref{eq:shah-lopatin-result}. The RHS shows the expressions derived from the Usadel equation technique as given in \eqref{eq:usadel-result-for-shah-lopatin}. Aligned adjacent cells or blocks of cells are (in total) equal.}
\label{tab:table}
\end{table*}

In this section, we summarize the most important aspects of a comparison between the results for the fluctuation conductivity obtained in this manuscript, see Sec.~\ref{subsec:genresults},  and those obtained with the help of the traditional diagrammatic technique in Ref.~\onlinecite{Lopatin05,Shah07}. The comparison is performed for a certain class of pair-breaking transitions considered in Ref.~\onlinecite{Lopatin05,Shah07}, for which our results are also applicable. As discussed in Sec.~\ref{subsec:genresults}, one- and two-dimensional systems in a parallel magnetic field can be treated on a similar footing. The relevant eigenfunctions of Eq.~\eqref{eq:eigenvalue-equation-1} are plane waves 
$\phi_{\bf q}(\mathbf{r})=e^{i\mathbf{qr}}$, where $\mathbf{q}$ is a one- or two-dimensional wave vector. The corresponding 
eigenvalues have the form $\alpha_q=\frac{1}{2}Dq^2+\alpha$, where $\alpha$ is the so-called pair-breaking 
parameter. In App.~\ref{app:appendix_shah} it is shown that the two sets of formulas for the fluctuation conductivity, one obtained from the diagrammatic calculation and the other one with the help of the Usadel equation approach, can be transformed into each other by means of algebraic manipulations [the ultraviolet cut-off in 2d and 3d requires a special discussion, see App.~\ref{app:appendix_shah}]. This is not only relevant from the point of view of consistency, but it enables us to find the relation between the classification used in Ref.~\onlinecite{Tikhonov12} as well as in this manuscript, and the traditional classification based on diagrams. 

In the diagrammatic technique, one deals with five core diagrams, displayed in the leftmost 
column of Tab.~\ref{tab:table}. Most of these have topologically similar partner diagrams, so that in total a standard set of 11 diagrams needs to be considered, see Ref.~\onlinecite{Larkin04}. In these diagrams, full lines represent quasiparticle propagators, while wavy lines represent fluctuation propagators. Disorder is taken into account with the help of the so-called impurity cross-technique,\cite{AGD63} shaded boxes and triangles, for example, symbolize Cooperons. 

The diagrams with one horizontal fluctuation propagator are referred to as density-of-states (DOS) diagrams, while those with vertical fluctuation propagators are called Maki-Thompson (MT) diagrams. Both of these diagrams may contain either two (2c) or three cooperons (3c). The final diagram is the so-called Aslamazov-Larkin (AL) diagram. It 
contains two fluctuation propagators. We would like to stress again that this classification is different from the one used in this manuscript. The authors of Ref.~\onlinecite{Lopatin05,Shah07} further distinguish terms for which the frequency integral contains the factor $\coth(\omega/2T)$ from those that contain $1/\sinh^2(\omega/2T)$. These are denoted as $ch$ and $sh$, respectively, in our notation they correspond to terms containing the bosonic distribution function $\mathcal{B}$ or its derivative $\mathcal{B}'$.
The terms $\delta\sigma^{(2c)}_{DOS,ch}$, $\delta\sigma^{(2c)}_{MT,ch}$ and 
$\delta\sigma^{(3c)}_{DOS,ch}$, $\delta\sigma^{(3c)}_{MT,ch}$ only differ in their overall coefficient
and in Ref.~\onlinecite{Lopatin05,Shah07} they are called $A$ and $B$, respectively. This naming scheme is also used here to refer to the terms obtained in Ref.~\onlinecite{Lopatin05,Shah07}, and they are displayed in the second column of Tab.~\ref{tab:table} next to the corresponding diagrams.
We now briefly describe the manipulations required to convert the equations
stated in Ref.~\onlinecite{Lopatin05,Shah07} to our results. More details can be found in Appendix~\ref{app:appendix_shah}.
The first step is to perform the integration over fermionic frequencies $\epsilon$ for all terms originating from the DOS- and MT-diagrams. The result of this integration can be expressed in terms of the propagator $\l$. After this step, 
the algebraic structure is similar to the corresponding terms in the technique used in this manuscript and first identifications can be made. Finally, it is necessary to 
integrate by parts in the radial momentum variable $q$. These manipulations lead from column two to column three in Tab.~\ref{tab:table}, and rearranging the resulting terms to obtain the 
results from the Usadel technique leads from column three to column four.

In summary, we were able to show that for a certain class of pair-breaking transitions there is a one-to-one correspondence between the results obtained from the diagrammatic perturbation theory and those obtained from the Usadel equation approach. The results of this comparison are illustrated in Tab.~\ref{tab:table}.

\section{Film in a tilted magnetic field}
\label{sec:tilted}

As an application of the formalism developed above, we calculate the fluctuation conductivity for a thin superconducting film 
penetrated by a magnetic field at an arbitrary angle. We focus on the low-temperature regime, where one can follow in detail the angular dependence of the resistance peak. The main results are summarized and discussed in Sec.~\ref{subsec:tilted}.  Here, we present details of the calculation.

First, one needs to solve Eq.~\eqref{eq:eigenvalue-equation-1} in order to find the eigenmodes and eigenvalues specific for the sample geometry and the orientation of the magnetic field. With this knowledge, the matrix elements $\rho_{nm}$ and ${\bf d}_{nm}$ of Eqs.~\eqref{eq:rho} and \eqref{eq:d} can be found. When inserted into the general expressions, Eqs.~\eqref{eq:dosansc} to \eqref{eq:general-al2b-1}, one can obtain a set of formulas that would in principle allow to evaluate the fluctuation corrections everywhere in the metallic phase. We will specialize on temperatures $T\ll T_{c0}$, however, where one finds comparatively simple results. In order to structure the calculation, it turns out to be useful to distinguish quantum contributions, that are independent of temperature, and persist for $T\rightarrow 0$, and thermal ones, which capture the temperature dependence, but vanish in the limit $T\rightarrow 0$. Two asymptotic regions in the phase diagrams are found where the evaluation of the sum over 
Landau levels can approximately be performed. As the lowest Landau level is of special importance in certain 
limits, we evaluate the contributions of the lowest Landau level and of the higher Landau levels separately.

\subsection{Eigenvalue equation and matrix elements}

Here, we solve the eigenvalue equation \eqref{eq:eigenvalue-equation-1} for a thin film in a tilted magnetic film. The tilting angle is denoted as $\theta$, for an illustration see Fig.~\ref{fig:A-sketch-of-the-system-1}. We describe the magnetic field by a vector potential in Landau gauge
\begin{equation}
A=\left(-\bperp y+\bpar z,0,0\right),
\end{equation}
where we abbreviated $B_{\perp}=B\sin\theta$, $B_{\parallel}=B\cos\theta$.
The eigenvalue equation \eqref{eq:eigenvalue-equation-1} then reads
\begin{equation}
-\frac{D}{2}\left(\left(\partial_{x}+2ie\bperp y-2ie\bpar z\right)^{2}+\partial_{y}^{2}+\partial_{z}^{2}\right)\phi=\alpha \phi.
\end{equation}

Under the assumption that the layer is thin, only the lowest
transverse mode in the $z$-direction is of interest, for which $\phi$ is almost constant as a
function of $z$. We drop $\partial_{z}^{2}$ and integrate over $z$ from
$-d/2$ to $d/2$:
\begin{align}
-\frac{D}{2}\left(\left(\partial_{x}-2ie\bperp y\right)^{2}+\partial_{y}^{2}\right)\phi & =\left(\alpha-\frac{D}{6}e^{2}\bpar^{2}d^{2}\right)\phi.
\end{align}
We thus find the eigenfunctions
\begin{equation}
\phi_{n p}=e^{ipy}\chi_{n}\left(x-\frac{p}{2e\bperp}\right),
\end{equation}
where $\chi_{n}$ are the eigenfunctions of a one-dimensional quantum
harmonic oscillator with frequency $2eD\bperp$ and mass $D^{-1}$,
and corresponding eigenvalues
\begin{equation}
\alpha_{n}=2eD\bperp\left(n+\frac{1}{2}\right)+\frac{D}{6}e^{2}\bpar^{2}d^{2}.\label{eq:alpha_n_landaulevel}
\end{equation}
The Landau levels are degenerate, as the eigenvalues do not depend on $p$. The perpendicular component of the magnetic field causes
Landau level quantization, and the parallel field component adds a constant to the pair-breaking parameter that is quadratic
in $\bpar$. It will prove useful to present the eigenvalues in the form
\be
\alpha_{n}=\alpha_{\perp}\left(2n+1+\mymu\right),
\ee
where $\mymu$ is introduced in Eq. \eqref{eq:mu-definition}.

The matrix elements required for the evaluation of the general formulas Eqs.~\eqref{eq:dosansc} to \eqref{eq:general-al2b-1} can
be found by using well-known relations for the eigenfunctions of a quantum
harmonic oscillator.\cite{Cohen80} As the eigenvalues do not depend on $p$, the integration over this variable can immediately be performed. We find
\begin{eqnarray}
	\int_{p,q} \rho_{np,mq} &=& \frac{B_{\perp}}{\pi} \delta_{nm}, \\
	\int_{p,q} \mathbf{d}^{x}_{np,mq}  &=&\frac{EB_{\perp}}{2\pi}\left(m\delta_{n+1,m}-n\delta_{n,m+1}\right).
\end{eqnarray}
For the second matrix element, we only calculated the vector component in the
$x$-direction, the direction of the electric field. This is sufficient for the calculation of the longitudinal conductivity. 

\subsection{Low temperature approximation, two asymptotic regimes}

In the low-temperature regime $t \ll 1$, the calculations can be considerably simplified by working with the asymptotic form of the fluctuation propagator. The calculation is further structured by separating quantum and thermal contributions as well as contributions of the lowest and of higher Landau levels. 

\subsubsection{Low temperature approximation and separation into thermal and quantum contributions}
For $t \ll 1$, we can use the asymptotic expansion for the digamma function and approximate the inverse fluctuation propagator, Eq.~\eqref{eq:E}, as
\begin{equation}
\e_{n}=-\ln\left(\frac{\alpha_{n}-\frac{i\omega}{2}}{\alpha_{c}}\right).\label{eq:low-temp}
\end{equation}

Considering further Eqs.~\eqref{eq:dosansc} to \eqref{eq:general-al2b-1}, we notice that terms with an integrand containing $\mathcal{B}'\left(\omega\right)=\partial_{\omega}\coth\left(\omega/2T\right)$ only contribute at finite temperatures as the integrand vanishes exponentially
as the temperature goes to zero. The terms containing $\mathcal{B}$, found from $\delta \sigma_{dos}$ and $\delta \sigma_{sc}^{\left(1\right)}$, contribute at finite temperatures as well as in the zero-temperature limit. 
For these terms, a useful separation can be achieved by writing 
\begin{equation}
	\mathcal{B}\left(\omega\right)=\left[\mathcal{B}\left(\omega\right)-\text{sgn}\left(\omega\right)\right]+\text{sgn}\left(\omega\right),
\end{equation}
where the combination in square brackets vanishes exponentially with temperature.
The notion quantum terms will be used for those contributions originating from $\text{sgn}\left(\omega\right)$ in this decomposition and they will be denoted as $\delta\sigma_0$; they are temperature-independent. 
The remaining terms, which encode the temperature-dependence and originate either from $\mathcal{B}'$ or from the difference $\mathcal{B}\left(\omega\right)-\text{sgn}\left(\omega\right)$, will be referred to as thermal terms and denoted as $\delta \sigma_T$; they vanish in the zero-temperature limit.

\subsubsection{Summation over Landau levels}

The evaluation of the formulas for the fluctuation corrections requires a summation over the Landau level index $n$ which in
general is difficult to treat.  However, we note that when approaching the
transition, $\l_{0}=1/\e_{0}$ diverges for $\omega=0$ while the other $\l_{n}$
remain finite. In order to quantify the importance of the $n=0$ term,
we introduce the parameter \begin{equation}
	\lambda=\frac{\l_{0}}{\l_{1}}-1=\frac{\ln\left(\frac{3+\mymu}{1+\mymu}\right)}{\ln
		\left(1+h\right)}.  
\end{equation} 
The number $\lambda$ measures the relative importance of the lowest Landau level as compared to
higher ones. The equations $\lambda \gg 1$ and $\lambda\ll 1$ define the
regions I and II, respectively, as introduced in Sec.~\ref{subsec:tilted}.

If $\lambda$ is large, the lowest Landau level is the only strongly
contributing mode. The divergence close to the critical line is accounted for by the term $n=0$ in the sum, a restriction to only this term is known as the
lowest Landau level approximation. This approximation has
been put forward by Galitski and Larkin \cite{Galitski01}
for the perpendicular magnetic field case $\mymu=0$, and allowed them to obtain closed formulas.

If $\lambda$ is small, the significance of the lowest Landau level is lost. To correctly understand the cross-over, we separate each contribution
in region I into four parts: First, each sum is split into the lowest Landau
level ($n=0$) term (LL), and the sum over higher Landau levels ($n>0$, HL).
Furthermore, it is of calculational advantage to perform a separation of each
of the resulting terms into the thermal (T) and quantum (0) contributions as explained above. 

\subsection{Evaluation of integrals in region I}

\subsubsection{Thermal terms}
\paragraph{Lowest Landau level (LL):}
For the evaluation of the thermal part, only frequencies $\omega\simeq T\ll T_{c0}$
contribute, and thus \eqref{eq:low-temp} can be expanded in $\omega$ and $h$. 
In order to find all relevant contributions, for the $dos$ and $an$ contributions
it is necessary to expand to second order.
The result of this expansion  can be written as 
\begin{equation}
\delta \sigma_{T,LL}=\frac{e^{2}}{\pi^{2}}\sum_{i}\int_{0}^{\infty}d\omega\frac{\alpha_{i}(\mathcal{B}-1)\omega-\beta_{i}\mathcal{B}'\omega^{2}}{\left(2\alpha_{0}h\right)^{2}+\omega^{2}},
\end{equation}
where $i$ enumerates different contributions. The prefactors $\alpha_{i}$
and $\beta_{i}$ are $\mymu$-dependent: \begin{equation}
	\begin{array}[c]{rclrcl}
	\alpha_{dos}	&=&-\frac{1}{1+\mymu}, &\beta_{dos}&=&-\frac{1}{1+\mymu}, \\
	\alpha_{an}	&=&0,&\beta_{an}&=&\frac{2}{1+\mymu}, \\
	\alpha_{sc}^{(1)}	&=&\frac{1}{3+\mymu},&\beta_{sc}^{\left(1\right)}&=&0, \\
	\alpha_{sc}^{\left(2a\right)}	&=&0,&\beta_{sc}^{\left(2a\right)}&=&\frac{1}{3+\mymu}, \\
	\alpha_{sc}^{\left(2b\right)}	&=&0,&\beta_{sc}^{\left(2b\right)}&=&\frac{4+2\mymu}{3+\mymu}.
\end{array}
\end{equation}
The integrals can be evaluated as a sum over poles of $\mathcal{B}$ and $\mathcal{B}'$. The result is the expression for $\delta \sigma_{T,LL}$ presented in Eq.~\eqref{eq:galitskii-larkin-general-alpha-beta}.

As compared to the result of Galitski and Larkin,\cite{Galitski01} we have modified
$\tilde{I}_\alpha$ by subtracting $\ln\left( h\right)$, which is part of the
quantum component [it will be treated separately below], and we added the
angular dependence via $\mymu$. Care must also be taken as Galitski and Larkin employed the traditional classification based on diagrams, while we use here the classification introduced in Ref.~\onlinecite{Tikhonov12}.

\paragraph{Higher landau levels (HL):} The quantum contribution due to the higher Landau levels are not singular at the transition and can be neglected. When going to smaller angles, the thermal contribution due to higher Landau levels increases. The point where the higher levels start to play a role marks the onset of the cross-over between region I and region II.

\subsubsection{Quantum terms}

\paragraph{Lowest Landau level (LL):}
The quantum part of the lowest Landau level contribution consists of terms originating from
$\delta \sigma_{dos}$ and $\delta \sigma_{sc}^{\left(1\right)}$. Contributions to the integral are not 
restricted to $\omega<T$ now, thus linearization of the integrand
is not appropriate here as it is for the thermal part.
In these terms, $\omega$ only appears in $\e$, thus we can make a change of variables 
$i\omega\rightarrow\tilde{\omega}$. The integration contour of $\tilde{\omega}$ can then be rotated by 90 degrees onto the real axis, thus making the integrand real.

Turning first to the quantum part of the DOS correction \eqref{eq:dos}, it  can be written as
\begin{equation}
\delta\sigma_{0,LL}^{\left(dos\right)}=-\frac{e^{2}}{2\pi^{2}}\left(\frac{1+h}{1+\mymu
}\right)\int_{1+h}^{\infty}\frac{dx}{x^{2}\ln x},
\end{equation}
and the integral can be expressed in terms of the logarithmic integral
function \cite{Abramowitz72} $\li\left(x\right)=\int_{0}^{x}\frac{dz}{\ln z}$, resulting in the first term in Eq.~\eqref{eq:sigma0LL}. 
The second contribution to the quantum part originates from $\sigma^{(1)}_{sc}$ of
Eq.~\eqref{eq:general-al1-1}. Upon setting $a=\frac{2(1+h)}{1+\mymu}$, it can be written as
\begin{equation}
\delta \sigma_{0,LL}^{\left(sc1\right)}=\frac{e^{2}}{2\pi^{2}}\int_{1+h}^{\infty}dx\left[\frac{1}{x}-\frac{1}{x+a}\right]\left[\frac{1}{\ln x}-\frac{1}{\ln x+a}\right].
\end{equation}
This gives the second term in Eq.~\eqref{eq:sigma0LL}. 
In order to approximate this integral, we note that in the limit $h\ll a$,
the asymptotic behavior at the phase boundary is due to the
singularity of the first inverse logarithm:
\begin{equation}
\delta \sigma_{0,LL}^{\left(sc1\right)}=\frac{e^{2}}{2\pi^{2}}\frac{a}{a+1}\ln h \qquad (h \ll a).
\end{equation}
Furthermore, the next term in this asymptotic expansion around small $h$ is an $a$-dependent
constant. For more general values of
$h$, the integral needs to be evaluated numerically.

\paragraph{Higher landau levels (HL):} The quantum contributions from $n>0$ are non-singular at the transition. Still, one needs to be more careful than for the thermal contribution, because the sum over all levels is in fact divergent, and a cut-off has to be introduced. 
We convert the sum over $n$ in the expressions \eqref{eq:dos} and \eqref{eq:general-al1-1} into an integral. This transformation becomes exact in
the limit $\theta\rightarrow0$, and it is a good approximation otherwise. 

The resulting integrals are doubly-logarithmically divergent, so a cut-off $\Lambda$
must be introduced, as discussed before Eq.~\eqref{eq:zero-temp-high-ll}.
After introducing dimensionless integration variables and again rotating the contour for $\omega$, we arrive at the integrals $\delta\sigma_{0,HL}^{(dos)}$ and $\delta\sigma_{0,HL}^{(sc1)}$ given in Eq.~\eqref{eq:zero-temp-high-ll}.

These integrals can also be expressed explicitely in terms of the logarithmic integral $\li(x)$. The correction then takes the form
\begin{eqnarray}
	\delta\sigma_{0,HL}=\frac{e^{2}}{2\pi^{2}}{\Bigg{[}}
	K \li\left(\frac{1}{K}\right) + K^2 \li \left(\frac{1}{K^{2}}\right)\no\\
	-\ln\frac{\ln K}{\ln a}
        +\mathcal{R}\left(a^2,K^2\right)-2\mathcal{R}\left(a,K\right){\Bigg]} ,
\label{eq:zero-temp-high-ll-as-li}
\end{eqnarray}
where we introduced the function
\begin{eqnarray}
\mathcal{R}(x,y)=x\left[\li\left(\frac{1}{x}\right)-\li\left(\frac{1}{y}\right)\right].
\end{eqnarray}
and abbreviated $a=(1+h)^2$ and $K=\Lambda/(2\alpha_{c0})$, and $\Lambda$ is the energy cut-off up to which superconducting fluctuations are taken into account. As discussed in Sec.~\ref{subsec:tilted}, 
it is of the order of the elastic scattering rate of electrons, $1/\tau$.

\subsection{Evaluation of integrals in region II}

For region II, the lowest Landau level loses its special significance. The
summation over Landau levels can be replaced by an integration. For the quantum
part $\delta\sigma_{0,HL}$, the result was already given in the previous chapter in Eq.~\eqref{eq:zero-temp-high-ll-as-li}.  We thus only need to
consider the thermal term $\delta\sigma_{T,HL}$ here.

In contrast to region I,
one needs to integrate over Landau levels also for the thermal contribution, an expansion of $\e$ in $\omega$ and
$h$, however, is again possible. 
After applying this expansion to all thermal terms, one finds that the term $\delta \sigma_{sc}^{(2b)}$ is dominant over all other terms, which are reduced by the factor $T/T_{c0}$, which in the considered regime is small. This is in contrast to the case of the $LL$, where all terms contribute with equal magnitude.

The dominant term $\delta\sigma_{sc}^{(2b)}$ can then be written as $\frac{4e^2}{\pi}F(\eta)$, 
with $F(\eta)$ given by an integral over $q$ and
 $\omega$,\cite{Shah07}
\begin{equation}
F\left(\eta\right)=\int\frac{d^{2}kdy}{\left(2\pi\right)^{2}}\frac{1}{
\sinh^{2}y}\frac{k^{2}y^{2}}{\left[\left(\eta+k^{2}\right)^{2}+y^
{2}\right]^{2}}.
\end{equation}
Here, $\eta$ is given by $\eta=(\alpha_{0}(h)-\alpha_{c})/T$, which for low temperatures can be written as 
$\eta=\frac{\pi}{2\gamma t}\left[2h+h^2\right]$.

The integration in $y$ can be performed
analytically after writing $1/\sinh^2 y$ as a sum of its poles, resulting in the following expression
\be
	&&F(\eta)\\
	&=&-\frac{1}{4 \pi^{2}}\int_{\eta/\pi}^{\infty}dx\ \left(\pi
	x-\eta\right)\left[\frac{1}{x}\psi'\left(x\right)+\psi''\left(x\right)+\frac{1}{2x^{3}}\right].\no
\ee
The integral $F$ can further be simplified to give Eq.~\eqref{eq:F}.
The asymptotic behavior of $F$ for large and small values of $\eta$ as stated below Eq.~\eqref{eq:Flimits} can be found
by inserting the asymptotic expansions $\psi'(x)\approx x^{-2}$ for small $x$
and $\psi'(x) = x^{-1} + x^{-2}/2 + x^{-3}/6 + \cdots$  for large $x$.

\section{Conclusion}
\label{sec:conclusion}

We studied the fluctuation conductivity of disordered superconductors subject to a magnetic field in the metallic phase using a quasiclassical kinetic equation approach. The derived expressions generalize the results of Ref.~\onlinecite{Tikhonov12}, in which calculations were performed for films in perpendicular magnetic fields, to a more general class of pair-breaking transitions in superconductors of different geometries. We were also able to make contact with previously derived  formulas for films and wires in parallel fields.\cite{Lopatin05, *Shah07} For the parallel magnetic field case we performed a detailed comparison  between the traditional classification based on diagrams with the classification based on the Usadel equation approach, see Tab.~\ref{tab:table}.  As an application, we studied fluctuation corrections in a superconducting film subject to a tilted field with emphasis on the low-temperature regime where these films display the phenomenon of the NM. The calculations performed in Ref.~\onlinecite{Tikhonov12} and in this manuscript clearly show in which way different physical mechanisms contribute to this phenomenon: The reduced quasiparticle density of states leads to an increase in resistance, while Cooper pairs do not efficiently transport charge in this regime and are thus unable to compensate this effect. 

It should be noted that besides the fluctuation corrections additional quantum corrections exist in disordered electronic systems. Both the weak localization correction and the Altshuler-Aronov interaction correction are logarithmically divergent at low temperatures. These corrections do not become singular near the phase transition, nevertheless they need to be taken into account in a quantitative comparison to experimental data. The dephasing time $\tau_\phi$ has been introduced phenomenologically in this paper, see Sec.~\ref{subsec:genresults}. For vanishing magnetic fields, dephasing is necessary in order to regularize the anomalous Maki-Thompson correction. At low temperatures and finite magnetic fields, no regularization is required, but  dephasing may still influence the magnetic field and temperature dependence of the fluctuation corrections. For example, it leads to a change in the phase boundary. 

The phenomenon of the NM is of particular relevance in the context of the magnetic field-tuned superconductor-insulator transition observed in thin disordered superconducting films.\cite{Hebard90} In these films, one finds a change from a superconducting to an insulating behavior ($d\rho/dT<0$) as a function of the magnetic field which leads to a very pronounced resistance maximum at low temperatures. For suitably prepared films, the resistance at the maximum may exceed the normal state resistance by many orders of magnitude.\cite{Sambandamurthy04} Unfortunately, transport in the highly resistive phase is very difficult to describe theoretically. The NM predicted\cite{Galitski01} and also observed\cite{Baturina05,Steiner06} in low-resistive films may be viewed as a precursor of this effect. Importantly, for low-resistive samples, controlled calculations can be performed, as was done in this manuscript. One may hope that a comparison to experimental data may contribute to a better understanding of the 
phenomenon of the NM and of the properties of the thin films in general.

\section*{Acknowledgments}

We are very grateful for discussions with A.~Finkel'stein, Ch.~Fr\"a\ss dorf, Y.~Oreg and K.~Tikhonov. We thank J.~Behrmann for help with the preparation of the manuscript. G.~S. acknowledges support by the Alexander von Humboldt Foundation.

\appendix

\section{Comparison to diagrammatic perturbation theory}
\label{app:appendix_shah}

In this appendix we show how the results of the diagrammatic calculation of Ref.~\onlinecite{Lopatin05, Shah07} and those obtained in the Usadel equation approach can be transformed into each other. The systems under study are films or wires subject to a parallel magnetic field. 

\subsection{Fluctuation corrections obtained from the Usadel equation technique}

In Ref.~\onlinecite{Lopatin05,Shah07}, the authors studied the fluctuation conductivity of 
systems for which the fluctuation spectrum is continuous in one or two dimensions and strongly discretized
in the transversal direction(s). For these systems, the eigenvalues of Eq.~\eqref{eq:eigenvalue-equation-1} are given by Eq.~\eqref{eq:alpha-pair-breaking-1}. The authors assume that only the lowest mode, $n=0$ is of
relevance. The corresponding eigenfunctions are plane waves in the unconfined directions, $\psi_{\bf q}=e^{i{\bf q}{\bf r}}$, where ${\bf q}$ is a one- or two-dimensional wave vector.

Under these assumptions, one can also obtain expressions for the fluctuation corrections from the general results presented in Sec.~\ref{subsec:genresults}.
Inserting the matrix elements of Eq.~\eqref{eq:rho} and Eq.~\eqref{eq:d}, restricting the sum over transverse modes to $n=0$ and writing $\alpha_{\perp 0}=\alpha$, the expressions for the fluctuation corrections can be written as
{\allowdisplaybreaks
\begin{align}
\delta\sigma_{dos} & =2De^2\int\frac{d\omega}{2\pi}\frac{d^{d}q}{\left(2\pi\right)^{d}}\left[\mathcal{B}'\,\mbox{Re}\e'\,\mbox{Im}\l-\mathcal{B}\,\mbox{Im}(\e^{\prime\prime}\l)\right],\nonumber\\
\delta\sigma_{an} & =2De^2\int\frac{d\omega}{2\pi}\frac{d^{d}p}{\left(2\pi\right)^{d}}\frac{\mathcal{B}'}{Dq^{2}/2+\alpha}\mbox{Im}\l\,\mbox{Im}\e,\nonumber \\
\delta\sigma_{sc}^{\left(1\right)} & =4D^{2}e^2\int\frac{d\omega}{2\pi}\frac{d^{d}q}{\left(2\pi\right)^{d}}q_{x}^{2}\,\mathcal{B}\,\mbox{Re}(\e''\e'\l^{2}),\nonumber\\
\delta\sigma_{sc}^{\left(2a\right)} & =-4D^{2}e^2\int\frac{d\omega}{2\pi}\frac{d^{d}q}{\left(2\pi\right)^{d}}q_{x}^{2}\,\mathcal{B}'\,\mbox{Re}\e^{\prime}\,\mbox{\ensuremath{\mbox{Re}}}(\l^{2}\e'),\nonumber \\
\delta\sigma_{sc}^{\left(2b\right)} & =16D^{2}e^2\int\frac{d\omega}{2\pi}\frac{d^{d}q}{\left(2\pi\right)^{d}}q_{x}^{2}\,\mathcal{B}'\,\mbox{Im}\e'\,\mbox{Re}(\e'\l)\,\mbox{Im}\l,\label{eq:usadel-result-for-shah-lopatin}
\end{align}}
where ${\bf q}$ is the
$d$-dimensional wave vector along the unconfined direction(s) ($d=1$
for wires, $d=2$ for films), we remind that $\mathcal{B}(\omega)=\coth(\omega/2T)$ is the bosonic distribution function, 
\be
&&\e_{q}\left(\omega\right) \no\\
&=&\ln\frac{T_{c0}}{T}+\psi\left[\frac{1}{2}\right]-\psi\left[\frac{1}{2}+\frac{Dq^2+2\alpha+i\omega}{4\pi T}\right],
\ee
and $\l=1/\e$ is the fluctuation propagator. In the displayed formulas the momentum and frequency arguments $(q,\omega)$ have been suppressed for the sake of brevity.

\subsection{Fluctuation corrections obtained in Ref.~\onlinecite{Shah07}}

We next compare the results \eqref{eq:usadel-result-for-shah-lopatin} 
to the results derived diagrammatically in Ref.~\onlinecite{Lopatin05,Shah07}. In that publication, the the following fluctuation corrections are presented [we adjusted them to the notation used in this manuscript]
{ \allowdisplaybreaks
\begin{widetext}
\begin{align}
\delta\sigma_{DOS}^{(sh,2c)} & =4iDe^{2}\int\frac{d^{d}qd\epsilon d\omega}{(2\pi)^{d}2\pi}\mathcal{H}\left(\epsilon\right)\mathcal{B}'\left(\omega\right)\l_{q}^{*}\left(\omega\right)\mbox{Re}\left[C_{q}^{2}\left(2\epsilon-\omega\right)\right],\no\\
\delta\sigma_{DOS}^{\left(sh,3c\right)} & =-8iDe^{2}\int\frac{d^{d}qd\epsilon d\omega}{(2\pi)^{d}2\pi}\mathcal{H}\left(\epsilon\right)\mathcal{B}'\left(\omega\right)\l_{q}^{*}\left(\omega\right)\mbox{Re}\left[Dq_{x}^{2}C_{q}^{3}(2\epsilon-\omega)\right],\nonumber \\
\delta\sigma_{MT}^{\left(sh\right)} & =4iD^{2}e^{2}\int\frac{d^{d}qd\epsilon d\omega}{\left(2\pi\right)^{d}{2\pi}}\mathcal{H}\left(\epsilon\right)\mathcal{B}'(\omega)\l^{*}_q(\omega)C_{q}\left(2\epsilon-\omega\right)C_{q}\left(-2\epsilon+\omega\right),\nonumber \\
-2A & =-16De^{2}\int\frac{d^{d}qd\epsilon d\omega}{\left(2\pi\right)^{d}2\pi}\mathcal{H}\left(\epsilon\right)\mathcal{B}\left(\omega\right)\left[C_{q}^{3}\left(2\epsilon-\omega\right)\l_{q}^{*}\left(\omega\right)\right],\nonumber \\
\frac{9}{2}B & =-72De^{2}\int\frac{d^{d}qd\epsilon d\omega}{\left(2\pi\right)^{d}2\pi}\mathcal{H}\left(\epsilon\right)\mathcal{B}\left(\omega\right)\l_{q}^{*}\left(\omega\right)Dq_{x}^{2}C_{q}^{4}\left(2\epsilon-\omega\right),\nonumber \\
\delta\sigma_{AL}^{\left(sh\right)} & =-8De^{2}\int\frac{d^dqd\omega}{(2\pi)^d2\pi}\mathcal{B}'\left(\omega\right)Dq_{x}^{2}\left[\left(\mbox{Re}[\l_q(\omega)\e^{\prime}_q(\omega)]\right)^{2}-\mbox{Im}\left[\l_q(\omega)\left(\e'_q(\omega)\right)^{2}\right]\mbox{Im}\l_q(\omega)\right],\nonumber \\
\delta\sigma_{AL}^{\left(cth\right)} & =16De^{2}\int\frac{d^dqd\omega}{(2\pi)^d2\pi}\mathcal{B}\left(\omega\right)Dq_{x}^{2}\,\mbox{Re}\left[\l^{2}_q(\omega)\e'_q(\omega)\e''_q(\omega)\right].\label{eq:shah-lopatin-result}
\end{align}
\end{widetext}
}
The corresponding diagrams are shown in Tab.~\ref{tab:table}. 

\subsection{Comparison}

The first five terms in Eq.~\eqref{eq:shah-lopatin-result} are not yet integrated with respect to the fermionic frequency $\epsilon$,
while the last two terms are already very similar in structure to 
$\delta\sigma_{sc}^{\left(1\right)},\ \delta\sigma_{sc}^{\left(2a\right)}$ and $\delta\sigma_{sc}^{\left(2b\right)}$
in \eqref{eq:usadel-result-for-shah-lopatin}. In fact, one can verify that
\begin{align}
4\delta\sigma_{sc}^{\left(1\right)} & =\delta\sigma_{AL}^{\left(cth\right)},\label{eq:sl-eq1-1}\\
2\delta\sigma_{sc}^{\left(2a\right)}+\delta\sigma_{sc}^{\left(2b\right)} & 
=\delta\sigma_{AL}^{\left(sh\right)}.\label{eq:sl-eq2-1}
\end{align}
The integration in $\epsilon$ for the first five terms in Eq.~\eqref{eq:shah-lopatin-result} can be
performed analytically. The required manipulations are similar to 
those described in App.~\ref{appendix_integration}. The results are
\begin{align}
\delta\sigma_{DOS}^{\left(sh,2c\right)} & =4e^2D\int\frac{d^dq d\omega}{(2\pi)^{d+1}}\mathcal{B}' \;\mbox{Im}\l\;\mbox{Re}\e^{\prime},\label{eq:shah-lopatin-integrated}\\
\delta\sigma_{DOS}^{\left(sh,3c\right)} & =-4e^2D\int\frac{d^dq d\omega}{(2\pi)^{d+1}}\mathcal{B}'Dq_{x}^{2}\;\mbox{Im}\l\;\mbox{Im}\e^{\prime\prime},\nonumber \\
-2A & =-8e^2D\int\frac{d^dq d\omega}{(2\pi)^{d+1}}\mathcal{B}\;\mbox{Im}\left[\l\e^{\prime\prime}\right],\nonumber \\
\frac{9}{2}B & =-12e^{2}D\int\frac{d^dq d\omega}{(2\pi)^{d+1}}\mathcal{B}\;Dq_{x}^{2}\;\mbox{Re}\left[\l\e^{\prime\prime\prime}\right],\nonumber \\
\delta\sigma_{MT}^{\left(sh\right)} & =2e^2D\int\frac{d^dq d\omega}{(2\pi)^{d+1}}\frac{1}{\alpha_{q}}\mathcal{B}'\;\mbox{Im}\l\;\mbox{Im}\e.\nonumber 
\end{align}

Comparing these expressions to \eqref{eq:usadel-result-for-shah-lopatin}, we find:
\begin{align}
2\delta\sigma_{dos} & =\delta\sigma_{DOS}^{\left(sh,2c\right)}-A,\label{eq:sl-eq3-1}\\
\delta\sigma_{an} & =\delta\sigma_{MT}^{\left(sh\right)}.\label{eq:sl-eq4-1}
\end{align}

Finally, by transforming the momentum integrals to spherical
coordinates and integrating by parts in the radial variable, one can
show that
\begin{align}
\frac{1}{2}\delta\sigma_{sc}^{\left(2a\right)} & =-\delta\sigma_{DOS}^{\left(sh,2c\right)}-\delta\sigma_{DOS}^{\left(sh,3c\right)},\label{eq:sl-eq5-1}\\
0 & =\frac{3}{2}A-3\delta\sigma_{sc}^{\left(1\right)}-\frac{9}{2}B.\label{eq:sl-eq6-1}
\end{align}

We note that for dimensions $d\ge2$ some of the terms in Eqs.~\eqref{eq:dosansc} to \eqref{eq:general-al2b-1} require an ultraviolet regularization. Indeed, an upper cut-off $\frac{1}{2}Dq^2\le\Lambda$ for the momentum integral needs to be introduced as discussed in connection with Eq.~\eqref{eq:zero-temp-high-ll}. In $2d$ this divergence is very weak (doubly logarithmic), in $3d$ it is more severe. This is not unusual as we work with a low-energy theory that ceases to be accurate at higher energies. The important point is that the theory captures correctly the sensitivity to temperature and magnetic fields. It should be remarked in this context that the boundary term, which we dropped when 
performing the integration by parts to obtain Eqs.~\eqref{eq:sl-eq5-1} and \eqref{eq:sl-eq6-1}, is in fact of order $\Lambda^{d-2}/\ln\Lambda$. The boundary term that arises once a finite cut-off is introduced, however, is insensitive to small changes in the parameters, $T$ and $h$. We therefore do not attribute particular importance to the difference in the ultraviolet regularization of the terms obtained from the Usadel equation and from the diagrammatic technique.

The identities \eqref{eq:sl-eq1-1}, \eqref{eq:sl-eq2-1}, \eqref{eq:sl-eq3-1},
\eqref{eq:sl-eq4-1}, \eqref{eq:sl-eq5-1} and \eqref{eq:sl-eq6-1}
show that our results, Eqs.~\eqref{eq:dosansc} to \eqref{eq:general-al2b-1}, when applied to the case under study, are equal to the corrections derived diagrammatically, Eq.~\eqref{eq:shah-lopatin-result}, in one spatial dimension. In $2d$ and $3d$, the equivalence still holds up to details of the ultraviolet regularization. The identification of corresponding terms is summarized in Tab. \ref{tab:table}.

\section{Frequency integration}
\label{appendix_integration}

The derivation of Eqs.~\eqref{eq:dosansc} to \eqref{eq:general-al2b-1} from Eqs.~\eqref{eq:jdos}, \eqref{eq:jan} and \eqref{eq:jsc} as well as the derivation of Eq.~\eqref{eq:shah-lopatin-integrated} from Eq.~\eqref{eq:shah-lopatin-result} both involve an integration over the ``fermionic'' frequency $\epsilon$. 
By way of example, we provide some details on the evaluation of the integrals. The integrand typically comprises two Cooperon propagators $C_n(\epsilon)=\left(i\epsilon-2\alpha_n\right)^{-1}$, multiplied by one or more factors of the fermionic equilibrium distribution function $\mathcal{H}_{\epsilon}=\tanh(\epsilon/2T)$ or its derivatives. 

Integrals with two Cooperons can be treated by a partial fraction decomposition. 
\begin{eqnarray}
A_{nm} &=& \int_\epsilon\; \mathcal{H}_{\epsilon}\;C_{n}\left(2\epsilon-\omega\right)C_{m}\left(2\epsilon-\omega\right)\nonumber \\
 & = & \frac{1}{2(\alpha_{n}-\alpha_{m})}\int_{\epsilon}\; \mathcal{H}_\epsilon\left[C_{n}\left(2\epsilon-\omega\right)-C_{m}\left(2\epsilon-\omega\right)\right]\nonumber \\
 & = & \frac{\e_{n}^{*}\left(\omega\right)-\e_{m}^{*}\left(\omega\right)}{4i\pi\left(\alpha_{n}-\alpha_{m}\right)},\label{eq:a_nm}
\end{eqnarray}
where we used the shorthand notation $\int_\epsilon=\int \frac{d\epsilon}{2 \pi}$.
The integral has been performed by the method of residues, the poles of $\mathcal{H}_\epsilon$ lie on the imaginary axis at $\epsilon_n = 4\pi i T \left(n+1/2\right),$ with $n$ integer. The resulting sum can be expressed in terms of the digamma function, and further in terms of $\e$, Eq.~\eqref{eq:E}. For $\alpha_{m}=\alpha_{n}$, one may use the relation $\partial_{\alpha_{n}}\e_{n}\left(\omega\right)=-2i\partial_{\omega}\e_{n}\left(\omega\right)$ and finds
\begin{equation}
A_{nn} = -\frac{1}{2\pi} \e^{*\prime}_{n}(\omega).
\end{equation}

Integrals involving $\mathcal{H}_{\epsilon-\omega}$ can be reduced to the discussed examples by a shift in the integration variable and complex conjugation, e.g., 
\begin{equation}
\int_{\epsilon}\mathcal{H}_{\epsilon-\omega}\;C_{n}^{2}\left(2\epsilon-\omega\right)=-A_{nn}^{*}.
\end{equation}
Another useful relation, that can be obtained in a similar way, is
\be
\bar{A}_{nm}  &=&  \int_\epsilon\; \mathcal{H}_{\epsilon}\; C_{n}\left(2\epsilon-\omega\right)C_{m}\left(-2\epsilon+\omega\right)\no \\
 &=&  \frac{\e_{n}^{*}\left(\omega\right)-\e_{m}\left(\omega\right)}{4i\pi\left(\alpha_{n}+\alpha_{m}\right)}.\label{eq:a_nm_bar}
\ee
Finally, integrals including $\mathcal{H}_\epsilon'$ can by reduced to the former integrals using integration by parts, yielding
\be
B_{nm}  &=&  \int_\epsilon\; \mathcal{H}_\epsilon'\;C_{n}\left(2\epsilon-\omega\right)C_{m}\left(2\epsilon-\omega\right)
  =  2A_{nm}^{\prime},\label{eq:b_nm}\\
\bar{B}_{nm}&=&\int_\epsilon\; \mathcal{H}_\epsilon'\;C_{n}\left(2\epsilon-\omega\right)C_{m}\left(-2\epsilon+\omega\right)=2\bar{A}_{nm}^{\prime}.\qquad\label{eq:b_nm_bar}
\ee

If the integrand contains more than one factor of $\mathcal{H}_{\epsilon}$ or $\mathcal{H}'_{\epsilon}$, the following two identities can be used to simplify the expression
\be
\mathcal{H}_\epsilon \mathcal{H}_{\epsilon-\omega}=1-\mathcal{B}_\omega\left(\mathcal{H}_\epsilon-\mathcal{H}_{\epsilon-\omega}\right),\label{eq:tanh-addition-theorem}\\
\mathcal{H}'_\epsilon \mathcal{H}_{\epsilon-\omega}=-\mathcal{B}_{\omega}\mathcal{H}'_{\epsilon}-\mathcal{B}'_\omega\left(\mathcal{H}_\epsilon-\mathcal{H}_{\epsilon-\omega}\right).\label{eq:tanh-addition-theorem-deriv}
\ee
The first identity is essentially the addition theorem for tanh \cite{Abramowitz72}, while second identity is obtained from the first one in three steps: A shift $\epsilon\rightarrow\epsilon+\omega$ is followed by a differentiation with respect to $\omega$ and a second shift $\epsilon\rightarrow\epsilon-\omega$.

With the stated integration formulas at hand, in combination with relations \eqref{eq:tanh-addition-theorem} and \eqref{eq:tanh-addition-theorem-deriv}, all the required integrals can be performed. Let us give an example, which is a step required for obtaining Eq.~\eqref{eq:dos} from Eq.~\eqref{eq:jdos}
\be
&&\int_{\epsilon}\;\mathcal{H}'_{\epsilon}\mathcal{H}_{\epsilon-\omega}\;C_{n}^{2}\left(2\epsilon-\omega\right)\no \\
&=& -\int_{\epsilon}\;C_{n}^{2}\left(2\epsilon-\omega\right)\left[\mathcal{H}'_{\epsilon}\mathcal{B}_{\omega}+\mathcal{B}'_{\omega}\mathcal{H}_{\epsilon}-\mathcal{B}'_{\omega}\mathcal{H}_{\epsilon-\omega}\right] \no\\
&=& -\mathcal{B}_{\omega}B_{nn}-\mathcal{B}'_{\omega}(A_{nn}+A_{nn}^{*}).
\ee


\end{document}